\documentclass[journal,11pt,onecolumn]{iopart}

\expandafter\let\csname equation*\endcsname\relax
\expandafter\let\csname endequation*\endcsname\relax
\usepackage{amsmath,amssymb}
\usepackage{amsthm,mathtools}
\usepackage{graphicx,bm,color,hyperref}
\usepackage[latin9]{inputenc}
\usepackage{cite} 

\DeclareMathOperator{\erf}{\mathrm{erf}}

\DeclareMathOperator{\var}{\mathrm{var}}

\DeclareMathOperator{\iu}{\mathrm{i}}
\DeclareMathOperator{\hiH}{\mathcal{H}}

\newtheorem{result}{Result}

\def\eff{\textrm{eff}}

\def\Gaps{\mathcal{G}}

\newcommand{\normone}[1]{\| #1 \|_1}

\newcommand{\ket}[1]{\vert #1 \rangle}
\newcommand{\bra}[1]{\langle #1 \vert}
\newcommand{\proj}[1]{\vert #1\rangle\!\langle#1 \vert}
\newcommand{\norm}[1]{\left\| #1 \right\|}
\newcommand{\av}[1]{\left\langle #1 \right\rangle}
\newcommand{\dd}{\textrm{d}} 

\newcommand{\teq}{T_{\textrm{eq}}} 

\definecolor{orange}{RGB}{200,66,0}
\definecolor{darkblue}{RGB}{0,0,139}

\begin{document}

\title{Equilibration time scales in closed many-body quantum systems}

\author{Thiago~R.~de~Oliveira$^{1}$, Christos~Charalambous$^2$, Daniel~Jonathan$^{1}$, Maciej~Lewenstein$^{2,3}$ and Arnau~Riera$^{2,4}$}

\address{$^1$Instituto de F\'isica, Universidade Federal Fluminense, Av. Gal. Milton
Tavares de Souza s/n, Gragoat\'a, 24210-346, Niter\'oi, RJ, Brazil}
\address{$^2$ICFO-Institut de Ciencies Fotoniques, The Barcelona Institute of Science and Technology, Castelldefels (Barcelona), 08860, Spain}
\address{$^3$ICREA-Instituci\'o Catalana de Recerca i Estudis Avan\c cats, Lluis Companys 23, Barcelona, 08010, Spain}
\address{$^4$ Max-Planck-Institut f\"ur Quantenoptik, Hans-Kopfermann-Str. 1, D-85748 Garching, Germany}

\begin{abstract}
We show that the physical mechanism for the equilibration of closed quantum systems is dephasing, and identify the energy scales that determine the equilibration timescale of a given observable. For realistic physical systems (e.g those with local Hamiltonians), our arguments imply timescales that do not increase with the system size, in contrast to previously known upper bounds. In particular we show that, for such Hamiltonians, the matrix representation of local observables in the energy basis is banded, and that this property is crucial in order to derive equilibration times that are non-negligible in macroscopic systems. Finally, we give an intuitive interpretation to recent theorems on equilibration time-scales.
\end{abstract}

\maketitle

\section{Introduction}

There is currently a renewed interest in the derivation of statistical mechanics from the kinematics and dynamics of a closed quantum system \cite{Gogolin16}. In this approach, instead of assuming \emph{a priori} that the system
is in some mixed state, such as e.g. a micro-canonical ensemble, one describes it at all times using a pure state $|\psi(t)\rangle$. One then seeks to show that, under reasonable conditions, the system behaves \emph{as if} it were described by a statistical ensemble. In this way the use of statistical mechanics can be justified without introducing additional external degrees of freedom, such as e.g. thermal `baths'.

A central part of this program has been to understand the process of \emph{equilibration}, i.e., how a constantly-evolving closed quantum system can behave as if relaxing to a stable equilibrium. The main insight relies on the fact \cite{Reimann2007,Goldstein2006,Popescu2006} that, if measurements are limited to small subsystems or restricted sets of observables, then `typical' pure states of large quantum systems are essentially indistinguishable from thermal states. It can then be shown \cite{Reimann2008,Linden2009} that under very general conditions on the Hamiltonian and nearly all initial states, the system will eventually equilibrate, in the sense that an (again, restricted) set of relevant physical quantities will remain most of the time very close to fixed, `equilibrium' values. 
For example, given some observable $A$ and a system of finite but arbitrarily large size, 
if its expectation value $\left\langle A(t)\right\rangle $ equilibrates,
then it must do so around the infinite time average (see Sec. 5.1 of \cite{Gogolin16})
\begin{equation}
\overline{A}=\lim_{T\rightarrow\infty}\frac{1}{T}\int_{0}^{T}\left\langle A(t)\right\rangle \,\,\dd t.
\end{equation}
If the infinite-time average fluctuation of $\left\langle A(t)\right\rangle $ around $\bar A$ is small, then we say that the observable $A$ equilibrates.

One major open question is to understand the time scale at which equilibration occurs in a given system, and in particular its scaling with respect to system parameters such as its size (number of degrees of freedom).
Various authors have tackled this question, e.g.
\cite{Monnai2013,Brandao2012,Cramer2012,Masanes2013,Malabarba2014,Goldstein2014,Goldstein2015,Garcia15,Short2012,Farrelly2016_QGases,Reimann2016,BalzReimann2017,ReviewSantos2017},
producing upper bounds that imply finite-time equilibration in various contexts (see also the `Supplementary Information' section in Ref. \cite{Reimann2016} for a brief survey of the literature).

Several of these results \cite{Brandao2012,Cramer2012,Masanes2013,Malabarba2014,Goldstein2014,Goldstein2015,Reimann2016,BalzReimann2017} are again obtained in a typicality framework: they estimate, in various different senses, the \emph{average} equilibration time of evolutions. While in many cases these calculated averages can have an impressive correspondence to experimentally measured equilibration times \cite{Reimann2016,BalzReimann2017}, this approach also has some inherent weaknesses. First of all, it is generally believed that many specific physical conditions that are realizable in Nature or in the lab can be very far from `typical' (for example, the actual Hamiltonians in Nature tend to have a locality structure that may be absent from most members of a mathematically generated ensemble \cite{BalzReimann2017}). In addition, by averaging, one loses information about the physical properties that are relevant to the equilibration time scale of any one specific evolution. 

Bounds on equilibration times without taking averages have also been obtained, but only for certain restricted classes of evolutions or observables, e.g. in Malabarba et al \cite{Malabarba2014}, Farrelly \cite{Farrelly2016_QGases}, Goldstein et al. \cite{Tasaki2013}, Monnai \cite{Monnai2013}, and Santos et al.\cite{ReviewSantos2017}.

Finally, there are a few works by the Bristol group \cite{Short2012,Garcia15} that derive general and rigorous bounds on the equilibration times of arbitrary observables, systems and initial states, without any ensemble averaging.  However, the bound in Ref.\cite{Short2012} scales with the inverse of the minimum energy difference (gap) in the system's spectrum. For physically realistic systems it  therefore increase exponentially with the system size, and cannot be a good estimate of the actual equilibration timescale - in particular since equilibration would then not occur in the thermodynamic limit (see section \ref{sec:discussion} for details). In contrast, a bound derived in Ref. \cite{Garcia15} can be independent of the system's size, however only in a regime that requires it to be initially in a nearly completely mixed state, failing to give a physically reasonable estimate in the case of a closed system in a pure initial state.

In this work, we seek to identify the properties of a closed quantum system which are relevant for the equilibration timescale of a given (arbitrary) observable.  Our approach is more heuristic than rigorous, but it allows us to estimate a timescale which, under reasonable circumstances, depends only weakly on the system size, and thus
seems to capture the relevant physics. The main insight we rely on is that equilibration is due primarily to a process of \emph{dephasing}  between different  Fourier components of the dynamical evolution -  a point that was briefly made in a classic reference \cite{Srednicki1994}, but that has apparently not been fully appreciated  by the current community. 

Although our argument does not result in a 
rigorous bound such as those in Refs. \cite{Garcia15,Short2012}, nor in a definite average evolution such as in Refs. \cite{Reimann2016,BalzReimann2017},
we are able to discuss how the equilibration timescale of a given observable $A$ depends on the physical properties of the system. Specifically, we find that the coherences of the observable 
of interest in the energy basis, $\langle E_{i}|A|E_{j}\rangle$, play a 
fundamental role. More specifically, the equilibration time depends critically on the range of energy gaps $E_{i}-E_{j}$ for which these coherences have non-negligible values. In particular, if this range remains roughly constant as the system size increases, the same will be true for the equilibration time. As we discuss below, this indeed happens for many observables of interest in many-body systems. We illustrate these results with numerical simulations of a spin chain, finding a reasonable qualitative agreement. It should be noted that similar heuristic methods and conclusions have also recently been proposed in simultaneous, independent research by Wilming et al. \cite{Wilming2017}.

This point of view also gives a new understanding of some existing results. For instance, it allows us to identify the reason for the limitations of bounds and estimates such as those in Refs. \cite{Short2012, Tasaki2013}, which, while rigorous, can vastly overestimate the time scale at which equilibration occurs in realistic systems.
In section \ref{sec:discussion}, we argue that the reason for this behaviour is that these estimates ultimately rely on 
the wrong physical mechanism of equilibration, disregarding the crucial role played by dephasing.

\noindent
Our main findings can be summarized as follows:
\begin{itemize}
\item In Sec.~\ref{sec:equilibration_as_dephasing} we discuss qualitatively why dephasing is the underlying mechanism of equilibration in closed quantum systems.
\item In Sec.~\ref{sec:Fourier} we develop a formalism based on the coarse-graining of functions in frequency space, which allows us to apply basic tools from Fourier transform theory, such as uncertainty relations, to equilibration related questions.
\item In Sec.~\ref{sec:energyscales} we determine the relevant energy scales that govern the equilibration time for a given observable, namely the energy fluctuations of the initial state and the bandwidth of the matrix of the observable in the energy basis. In particular, we give an independent proof of the fact \cite{Arad2016} that local observables have banded matrices when written in the energy basis of a short-ranged spin Hamiltonian.
\item In Sec.~\ref{sec:XXZmodel} we illustrate our results with a numerical simulation of the XXZ model.
\item In Sec.~\ref{sec:discussion} we discuss some implications of our results. In Sec.~\ref{subsec:reinterp}, we reinterpret existing results from the point of view of our dephasing framework. In  Sec. \ref{subsec:levelstat} we discuss implications for the fields of quantum chaos and integrability. In particular we present two models with identical eigenbases but different level statistics that have indistinguishable dynamics over realistic time-scales.
\end{itemize}

\section{General setting and definition of the problem}

Let us consider a closed system whose state is described by a vector
in a Hilbert space of dimension $d_{T}$ and whose Hamiltonian has
a spectral representation 
\begin{equation}
H=\sum_{k=1}^{d_{E}}E_{k}P_{k}\,,
\end{equation}
where $E_{k}$ are its energies and $P_{k}$ the projectors onto its
eigenspaces. Note that the sum runs over $d_{E}\le d_{T}$ terms,
since some eigenspaces can be degenerate.

We denote the initial state by $\ket{\psi(0)}$. If the Hamiltonian
has degenerate energies, we choose an eigenbasis of $H$ such that
$\ket{\psi(0)}$ has non-zero overlap with only one eigenstate $\ket{E_{k}}$
for each distinct energy. Choosing units such that $\hbar=1$, the
state at time $t$ is then given by 
\begin{equation}
\ket{\psi(t)}=\sum_{k}c_{k}e^{-iE_{k}t}\ket{E_{k}},
\end{equation}
with $c_{k}\equiv\bra{E_{k}}\psi(0)\rangle$. It is clear that $\ket{\psi(t)}$
evolves in the subspace spanned by $\{\ket{E_{k}}\}$ as if it were
acted on by the non-degenerate Hamiltonian $H'=\sum_{k}E_{k}\proj{E_{k}}$.
In this case, if the system equilibrates, the equilibrium state must
be
\begin{equation}
\omega=\sum_{k}|c_{k}|^{2}\proj{E_{k}}\,.
\end{equation}

In this article, following a number of authors, \cite{Reimann2008,Linden2009,Short2012,Garcia15}
we will study equilibration by focusing on observables. The idea is
that a system can be considered in equilibrium if all experimentally
relevant (typically, coarse-grained) observables $A$ have equilibrated.
In other words, we will focus on understanding how the expectation
value $\left\langle A(t)\right\rangle $ approaches its equilibrium
value $\Tr(A\omega)$. Note that, in order to even talk about equilibration
time scales, we must assume that such a condition holds, i.~e., that
this observable sooner or later equilibrates.

Let us introduce the \emph{time signal} of $A$, given the initial
state $\ket{\psi(0)}$, as the distance of $\av{A(t)}$ from equilibrium at
time $t$ 
\begin{equation}
 \begin{split}
g(t)\coloneqq & \;\frac{1}{\Delta_A}\left(\bra{\psi(t)}A\ket{\psi(t)}-\Tr(A\omega)\right)\\
= & \frac{1}{\Delta_A}\sum_{i\ne j}(c_{j}^{*}A_{ji}c_{i})\e^{-\iu(E_{i}-E_{j})t}\, , \label{eq:timesignal}
\end{split}
\end{equation}
where $A_{ij}\coloneqq\bra{E_{i}}A\ket{E_{j}}$ are the matrix elements of
$A$ in the energy eigenbasis, and $\Delta_A=a_{\max} -a_{\min}$ is the range of possible outcomes, 
being $a_{\max (\min)}$ the largest (smallest) eigenvalue of $A$. 
The denominator $\Delta_A$ is
introduced to make the time signal dimensionless and satisfying $ | g(t) | \leq 1$.
 Note that the time
signals of two observables $A$ and $A'=b(A-a_0)$ are identical for $a_0, b\in\mathbb{R}$. 

We can conveniently rewrite the time signal as
\begin{equation}
g(t)=\sum_{\alpha\in\mathcal{G}}v_{\alpha}\e^{\iu G_{\alpha}t},\label{eq:thesum}
\end{equation}
where $\alpha \in \Gaps=\{(i,j):i,j\in\{1,\ldots d_{E}\},i\ne j\}$ labels each energy gap $G_{\alpha}=(E_{j}-E_{i})$ appearing in the system's spectrum, and where 
\begin{equation} \label{eq:v-alpha}
v_{\alpha}=v_{(i,j)}=\frac{c_{j}^{*}A_{ji}c_{i}}{\Delta_A }.
\end{equation}
We will refer to the complex number $v_{\alpha}$ as the \emph{amplitude} of the corresponding gap $G_{\alpha}$, and to its normalized square modulus $	q_{\alpha}\coloneqq |v_{\alpha}|^{2} / \sum_{\beta} |v_{\beta}|^{2}$ as the \emph{relevance} of $G_{\alpha}$. Note that the set of relevances form a probability distribution over $\mathcal{G}$.

A physical interpretation of this normalization factor can be given as follows \cite{Reimann2008,Linden2009,Short2011}: note that, if the system has \emph{non-degenerate gaps}, then the time-averaged fluctuations of the time signal
\begin{equation}
\av{|g|^{2}}_{T}:= \frac{1}{T}\int_{0}^{T}\dd t|g(t)|^{2}\label{eq:finite-time-average-distance}
\end{equation}
satisfy the limit
\begin{equation}
 \av{|g|^{2}}_{\infty}:= \lim_{T\to\infty}\av{|g|^{2}}_{T}=\sum_{\alpha}|v_{\alpha}|^{2}.\label{eq:time-average-distance2}
\end{equation}
In other words, this quantity gives  the infinite time average
of (the square of) the deviation of $\av{A(t)}$ from its equilibrium value. We consider that the observable $A$ equilibrates if this quantity is small, in the sense that
\begin{equation}
\av{|g|^{2}}_{\infty} \lesssim 
\left(\frac{\delta A}{\Delta_A}\right)^{2}\,,\label{eq:equilibration_definition}
\end{equation}
where $\delta A$ is the experimentally available resolution. Furthermore, note that $\Delta_A/\delta A$ quantifies the
the amount of different possible outcomes of a measurement and hence 
$\delta A/\Delta_A$ is expected to be much much smaller than 1.
From now on, without loss of generality and for the sake of simplicity we will consider that
$\Delta_A=1$.

In Refs.~\cite{Reimann2008,Linden2009,Short2011}, sufficient conditions
for equilibration in the sense defined above are given by bounding the average distance from
equilibrium \eqref{eq:time-average-distance2}. In particular, in
\cite{Short2011} it is shown that for any Hamiltonian with non-degenerate
gaps, 
\begin{equation}
\sum_{\alpha}|v_{\alpha}|^{2}\le\frac{1}{d_{\eff}},\label{eq:Short-bound}
\end{equation}
where the effective dimension is defined as $d_{\eff}:=1/(\sum_{k}|c_{k}|^{4})=1/{\Tr(\omega^{2})}$,
which roughly speaking tells us how many eigenstates of the Hamiltonian
participate in the superposition of the initial state. Thus, a large
effective dimension, which is usually the case in many body systems
 (see \ref{sec:local-Ham}
and Ref.~\cite{Reimann2010}), 
is sufficient to guarantee that condition (\ref{eq:equilibration_definition})
will be satisfied. Concerning the assumption of the Hamiltonian having
non-degenerate gaps, it is shown in \cite{Short2012} that as long
as there are not exponentially many degeneracies the argument stays
the same.

\section{Equilibration as dephasing}
\label{sec:equilibration_as_dephasing}
Let us consider a situation in which the initial state is out of equilibrium,
i.~e.\ the time signal of a given operator is initially significantly
larger than the equilibrium value: $|g(0)|\gg \sqrt{\av{|g|^{2}}_{\infty}}$. For this to happen, 
the \emph{phases of the complex numbers $v_{\alpha}$ in the time signal \eqref{eq:thesum}
need to be highly synchronized}.  This case is presented pictorially in Fig.~1 (left) where the $v_{\alpha}$'s are depicted as points in the complex plane.

\begin{result}[Equilibration is dephasing] Given a time signal
$g(t)=\sum_{\alpha}v_{\alpha}\e^{\iu G_{\alpha}t}$ with $v_{\alpha}=|v_{\alpha}|\e^{\iu \theta_\alpha}\in\mathbb{C}$
being the initial amplitude of the gap $G_{\alpha}\in\mathbb{R}$, 
a necessary condition for the system to be initially out of equilibrium, i.e., $|g(0)|$ significantly
larger than the typical equilibrium fluctuation $\sqrt{\av{|g|^{2}}_{\infty}}$,
is that the initial phases $\theta_{\alpha}$ are not isotropically distributed but
significantly synchronized.
More precisely, we quantify the distance from equilibrium as
\begin{equation}\label{eq:synchronized-phases}
|g(0)|^2-\av{|g|^2}_\infty = 2\sum_{\alpha < \beta} |v_\alpha||v_\beta|\cos(\theta_\alpha - \theta_\beta)\, ,
\end{equation}
which becomes negligible when the phases $\theta_\alpha$ are isotropically distributed.
\end{result} 

Equation \eqref{eq:synchronized-phases} follows from a straightforward calculation
\begin{equation}
|g(0)|^2=\left(\sum_\alpha v_\alpha \right)\left(\sum_\beta v_\beta \right)^*=\av{|g|^2}_\infty + 2\sum_{\alpha < \beta} |v_\alpha||v_\beta|\cos(\theta_\alpha - \theta_\beta)\, .
\end{equation}
To see that isotropic randomly distributed
phases give $|g(0)|^{2}\simeq \av{|g|^{2}}_{\infty}$,
let $v_{\alpha}=|v_{\alpha}|\e^{\iu\theta_{\alpha}}$ 
be a set of independent random complex variables with an isotropic
probability distribution $p_{\alpha}(r,\theta)=p_{\alpha}(r)=\frac{1}{2\pi}\delta(r-r_{\alpha})$,
i.~e.\ the random variable $v_{\alpha}$ has fixed modulus $r_{\alpha}$
and a random phase $\theta_{\alpha}$ uniformly distributed around the circle. 
Then, the variance of the random
variable $\sum_{\alpha}v_{\alpha}$ is 
\begin{equation}
\var\left(\sum_{\alpha}v_{\alpha}\right)=\sum_{\alpha}\var(v_{\alpha})=\sum_{\alpha}\left\langle |v_{\alpha}|^{2}\right\rangle =\sum_{\alpha}|v_{\alpha}|^{2}
\end{equation}
where we have used the fact that the variance of a sum of independent
random variables is the sum of variances and the first moments $<v_{\alpha}>=0$.
That is, if the phases of $v_{\alpha}$ are uniformly distributed, then the typical
initial value of the time signal is $|g(0)|=|\sum_{\alpha}v_{\alpha}|\simeq\left(\sum_{\alpha}|v_{\alpha}|^{2}\right)^{1/2}$.

\begin{figure}
\vspace{1cm}
 \begin{center}
 \begin{tabular}{cc}
\includegraphics[width=0.47\columnwidth]{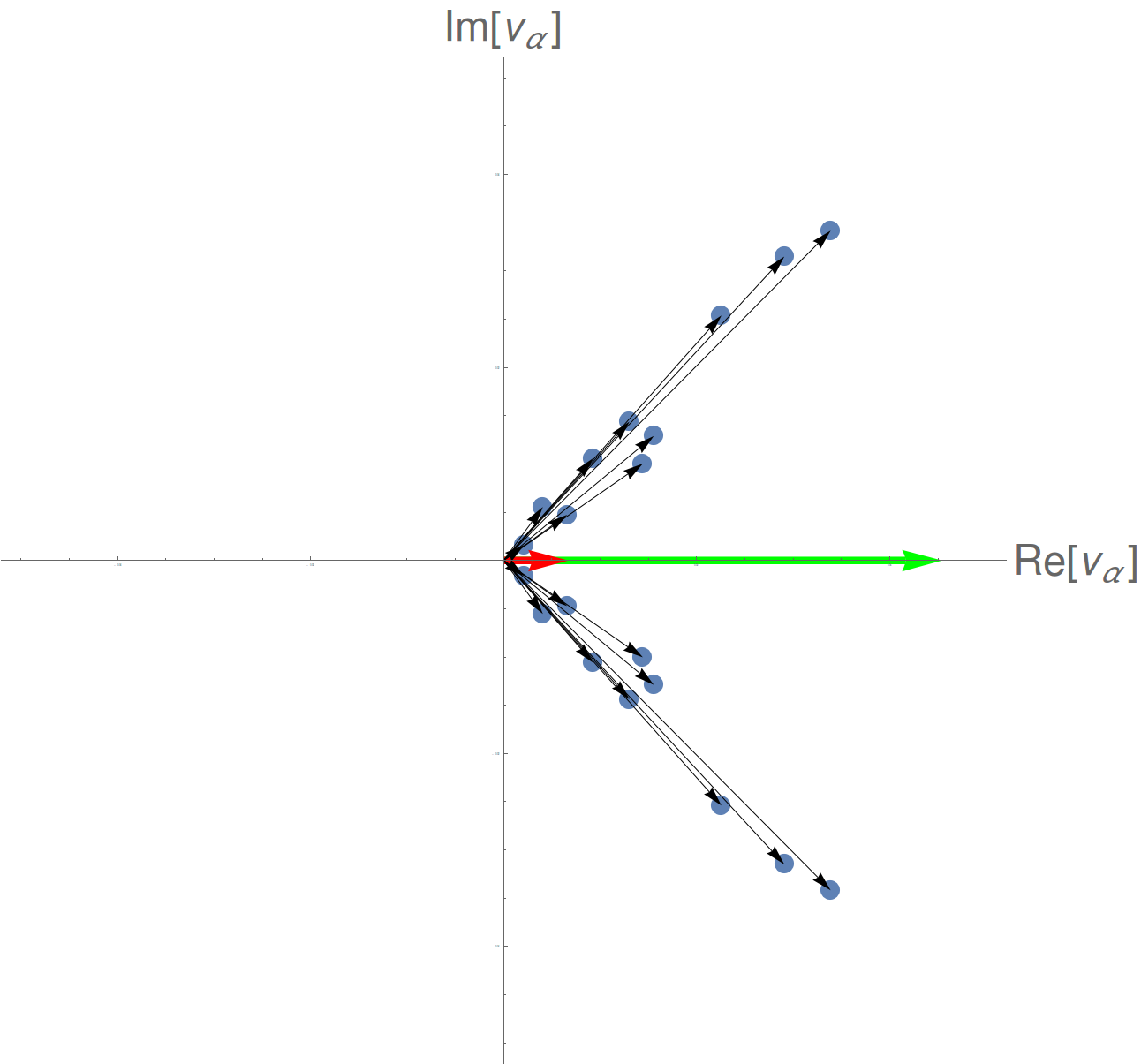}  & \includegraphics[width=0.47\columnwidth]{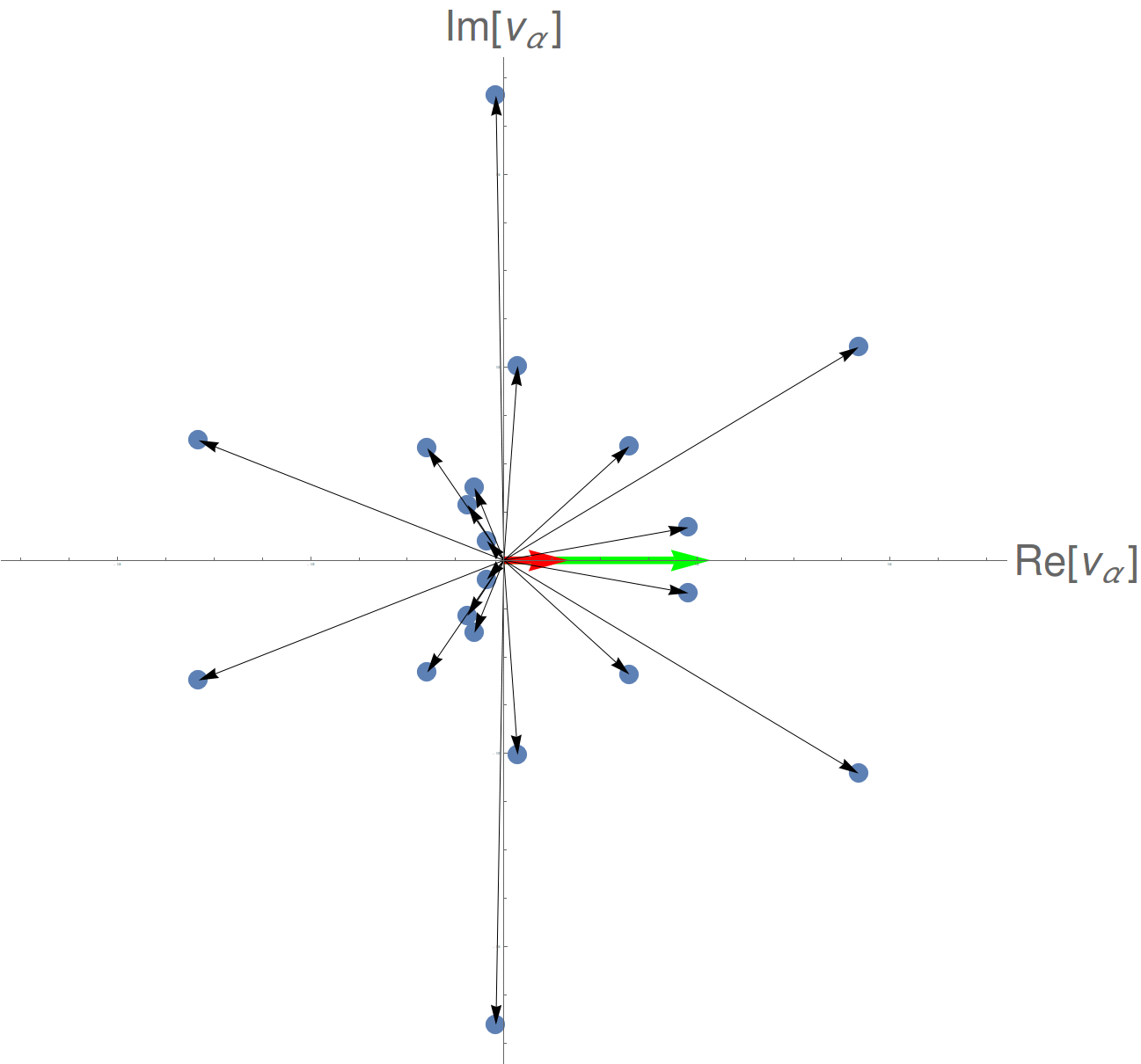} \tabularnewline
\end{tabular}
 \end{center}
\caption{ \label{fig1}
(Color online). Illustration of the dephasing process of the complex terms $v_\alpha \e^{\iu G_\alpha t}$ (blue dots/black arrows) of a time signal $g(t)$ (green arrow), see Eq.(\ref{eq:thesum}). On the left, the system is far from equilibrium, a short time $t=0$, having been initialized with all $v_{\alpha}$ real, and $g(t)$ is substantial. Note that half of the complex terms have rotated clockwise, and the other half anti-clockwise, as expected from the symmetry of the set of gaps $G_{\alpha}$.  On the right, after a long time, 
the individual complex vectors have become spread out, and the system has equilibrated, with $g(t)$ becoming close to the typical fluctuation $\sqrt{\av{|g|^{2}}_{\infty}}$ (red arrow).
}
\end{figure}

In contrast, an out-of-equilibrium initial state, for a distribution of gaps with a non-zero dispersion and a large enough system, will evolve to an equilibrium state, represented by an isotropic cloud of points centered at the origin of the complex plane as shown in Fig.~1 (right). This mechanism for equilibration, or more generally the vanishing amplitude of a signal over time, is usually called \emph{dephasing} and is a well-known feature in many different fields of physics \cite{Fonda,Robinett2004,Eberly1980, Narozhnyetal,AllenEberlyBook,PitaevskiiBook, BaymPethickBook,ReviewSantos2017}, 

The time-scale necessary for the sum in Eq.(\ref{eq:thesum}) to dephase can be estimated with the following simple argument. 
Suppose for simplicity that all terms have the same phase at $t =0$. One way to lower bound the time needed for these phases to
spread around the whole range $[0,2\pi)$ is to use the time it takes
for the difference between the fastest and slowest phase to differ
by $2\pi$: $tG_{\max}-tG_{\min}=2\pi$ which leads to $T_{eq}\sim2\pi/(G_{\max}-G_{\min})$.
However, the slowest and fastest gaps are not necessarily very relevant to the sum, that is, they
can have a relatively small amplitude $v_{\alpha}$.
For this reason, a better estimate is obtained by replacing the denominator $G_{\max}-G_{\min}$ with $2 \sigma_{G}$, where we define the \emph{gap dispersion}  $ \sigma_{G}$ to be simply the standard deviation of the  $G_{\alpha}$
when weighted by their respective relevances $q_{\alpha}$, i.~e. 
\begin{equation}\label{eq:gap-dispersion}
\sigma_{G}^{2}:=\sum_{\alpha}q_{\alpha}(G_{\alpha}-\mu_{G})^{2}.
\end{equation}
Since for each gap $G_{(i,j)}=E_{j}-E_{i}$ there is
also $G_{(j,i)}= -G_{(i,j)}$, then  by symmetry  the average gap is $\mu_{G} =0$, and also $|v_{(i,j)}|=|v_{(j,i)}|$.
By the argument made above, we can thus expect that, at least in cases
where the distribution $G_{\alpha}$ is of a unimodal type, the equilibration
time can be estimated by 
\begin{align}\label{eq:Teq}
\teq\sim \pi/\sigma_G. 
\end{align}
Note that similar estimates are made in the various fields where dephasing is relevant, e.g.  in \cite{Narozhnyetal,PitaevskiiBook, BaymPethickBook}. In the next section we will give a more detailed justification for this estimate using standard tools from Fourier transform theory. 

The physical properties that control the dispersion $\sigma_G$, and thus the equilibration time $\teq$, can be identified once we note that the probabilities $q_\alpha$ are proportional to the $|v_{\alpha}|^2$. Since $v_{\alpha}\propto c_{i}^{*}c_{j}\langle E_{i}|A|E_{j}\rangle$ (Eq. (\ref{eq:v-alpha})), then $\sigma_G$ is determined by: 
i) the probability distribution $|c_{i}|^2$ for the energies;
ii) the matrix-elements of the observable $A$ in the energy basis;  and
iii) the distribution of the values of the gaps $G_\alpha=E_j-E_i$ themselves.

In Sec.~\ref{sec:energyscales} we identify the energy scales of the observable and the initial state which are relevant to determine 
the dispersion of gaps $\sigma_G$.
Let us now anticipate the requirements on the initial state and the observable in order that $\sigma_G$ does not diverge in the thermodynamic limit, i.e. when the system size $n\to\infty$ and the equilibration time $\teq$ does not vanish. We can identify two regimes where this will happen: 

\begin{itemize}
\item First, there are observables $A$ that equilibrate in finite time regardless of the details of the initial state. Note that, for large many-body systems, the variance of the density of states  usually scales as $\sqrt{n}$, and for generic states the width of the distribution of $|c_{i}|^{2}$ is as wide as the energy spectrum, which increases with $n$. This implies that, unless restrictions are placed on $A$, the gap variance $\sigma_G$ will have the same scaling and $\teq$ will vanish in the thermodynamic limit. In order to avoid this problem, the variance of the distribution  of $v_{\alpha}$ should not increase with $n$. This requires that the distribution $\langle E_{i}|A|E_{j}\rangle$ should decrease (or become null) for large values of $E_{j}-E_{i}$ - in other words, $A$ needs to be \emph{banded} in the energy basis.

\item Conversely, suppose the initial state has support over an energy range that does not scale with $n$. This can happen, for example, in the case of a so-called `local quench' \cite{Polkovnikov11}, when a local subsystem of fixed dimension is excited regardless of the full size of the system. In this case, $\sigma_G$  will also at worst be independent of $n$, and so even observables with long-range coherence between very different energies, which would otherwise equilibrate quickly, will now take a finite time $\teq$. 
\end{itemize}
Finally, it is also important to mention that the gap dispersion
 $\sigma_G$ may actually decay to zero with $n$, leading to equilibration times that become very long. 
As an example of this situation consider 
two subsystems of increasing size $n$ interacting through a spatially localized border of fixed size. The coherence in the energy basis of the interaction Hamiltonian is bounded by the operator norm of such an interaction, matching with the intuition that the stronger the interaction between systems, the faster the relaxation, and vice-versa. By rescaling the global Hamiltonian, we can see that the interaction terms become relatively weaker as $n$ grows and thereby the equilibration slower.

\section{Fourier description of the dephasing framework} \label{sec:Fourier}

In this section we give further substance to the above heuristic argument to estimate the equilibration time-scale by means of Fourier transform techniques. 

\begin{figure}
\begin{center}
\includegraphics[width=.8\textwidth ]{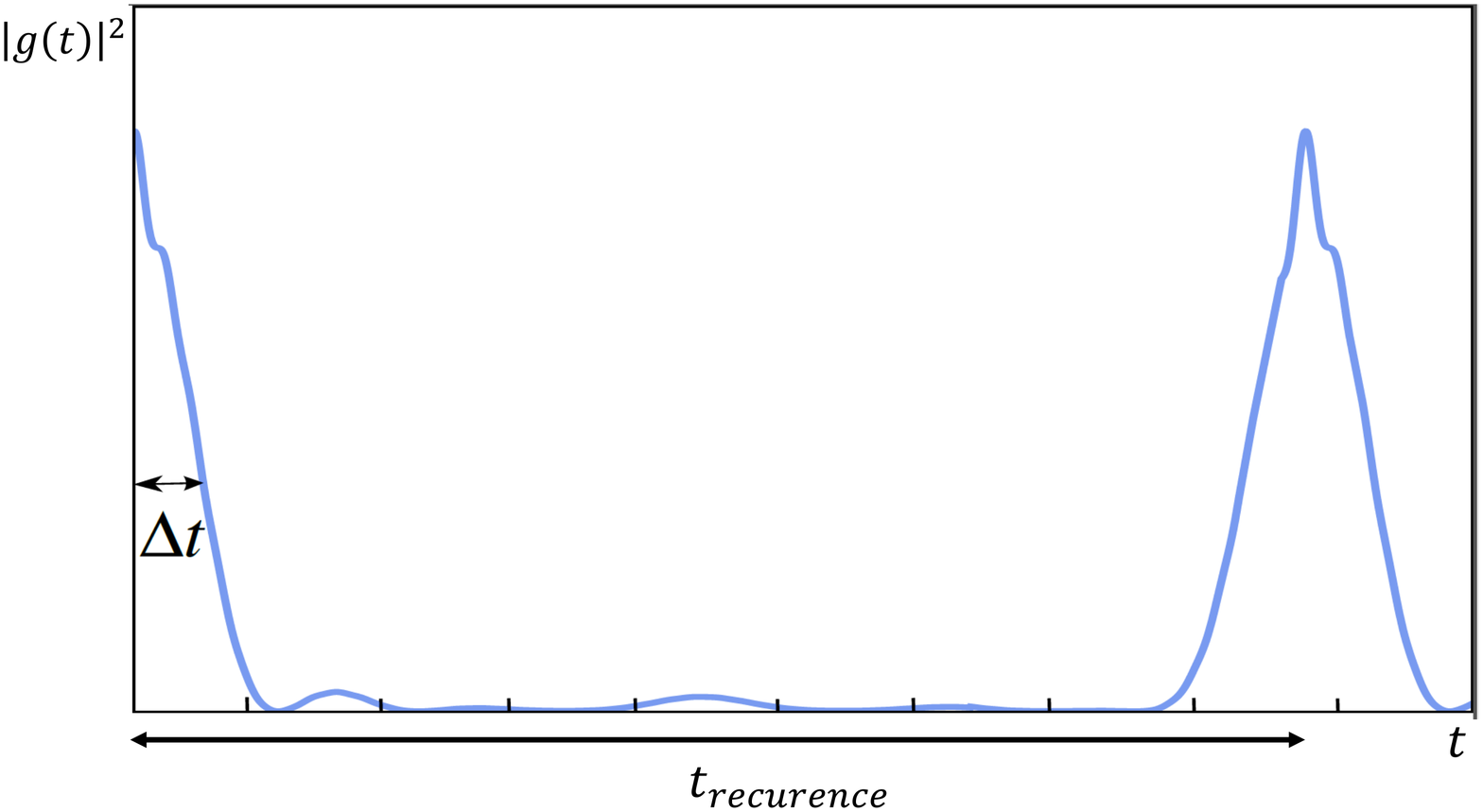}
\end{center}
\caption{Example of a time signal which has a recurrence time. 
In the absence of recurrences, a good estimate the equilibration time scale $T_{eq}$ would be a few multiples of $\Delta t$ defined in Eq.~\eqref{eq:time-fluctuations}.
\label{fig:time-signal}
}
\end{figure}

Let us first give a general idea of our approach. Suppose the time-signal $g(t)$ decayed more or less steadily to zero, and stayed there. If so, then a good estimate for the equilibration time scale would be given by a few multiples of the standard deviation $\Delta t$ defined by
\begin{equation}\label{eq:time-fluctuations}
\Delta t^{2}\coloneqq \frac{1}{\norm{g}_{2}}\int\dd t |g(t)|^{2}(t-\mu_{t})^{2}
\end{equation}
with $\mu_{t}\coloneqq \int\dd t|g(t)|^{2}t/\norm{g}_{2}$ (see Fig.~\ref{fig:time-signal}). 

In such a case, following the spirit of our previous heuristic example, it would also be tempting to estimate the order of magnitude of $\Delta t$ by taking the inverse of 
the spectral variance $\Delta \omega$ of the signal. Indeed, this can be justified if we recall the standard uncertainty principle of Fourier analysis \cite{Stein2003} 
\begin{equation} \label{eq:standardUP}
\Delta t\cdot\Delta\omega\ge1/2.
\end{equation} 
Of course, this is a lower bound, that is saturated exactly only in the case of a Gaussian spectrum. However, it will be nearly saturated ($\Delta t\cdot\Delta\omega = c_{1}$ , where $c_{1}$ is a constant of order 1), when the spectrum is unimodal and without long tails. In this case, we can also expect the time signal to decrease to a very small value after a time  $c_{2} \Delta t$, for some small multiple $c_{2}$ again of order 1. Taking both multiples together, we expect a good estimate for the equilibration time to be of order $\sim \frac{c_{1}c_{2}}{ 2 \Delta\omega}$.This is satisfied by  Eq. (\ref{eq:Teq}), which we will therefore continue to take as our estimate. 

Unfortunately, for finite systems, our initial assumption of steady decay does not apply. The time signal has recurrences, that is, a long time after the dephasing has occurred and the system has equilibrated, the phases get again aligned (synchronized) in the complex plane, and the signal regains strength (Fig.~\ref{fig:time-signal}). In order to avoid this problem, in the next subsections we introduce a coarse-grained version of the signal spectrum, which dampens out the recurrences. This allows us to exploit
the uncertainty principle to estimate the equilibration time-scales, as described above.
Under some mild conditions, we show that the equilibration time-scale estimated provided by this procedure coincides with the one previously given by the heuristic argument of points dephasing in the complex plane.

\subsection{The frequency signal}

We define the \emph{frequency signal}, $\tilde{g}(\omega)$, as the Fourier transform of the time signal
$g(t)$ 
\begin{equation}\label{eq:freq-signal}
\tilde{g}(\omega)\coloneqq \mathcal{F}[g](\omega) = \frac{1}{\sqrt{2\pi}} \int_{-\infty}^{\infty}g(t)e^{-i\omega t} dt\,.
\end{equation}
which roughly speaking tells us the relevance with which every frequency
contributes to the time signal. When both the time and frequency signals are square-integrable ($g,\tilde{g}\in L^{2}$), the standard uncertainty principle of Eq. (\ref{eq:standardUP}) applies, where
\begin{equation} \label{eq:deltaomega}
\Delta\omega^{2}\coloneqq \frac{1}{\norm{\tilde{g}}_{2}}\int\dd\omega|\tilde{g}(\omega)|^{2}(\omega-\mu_{\omega})^{2}
\end{equation}
with $\mu_{\omega}\coloneqq \int\dd\omega|\tilde{g}(\omega)|^{2}\omega/\norm{\tilde{g}}_{2}$.

However, in the case of time signals such as in Eq. (\ref{eq:thesum}),  $g(t) \not\in L^{2}$, as can be seen from Eq. (\ref{eq:time-average-distance2}) (the integral diverges proportional to $T$). 
The same is true for the frequency signal, since
\begin{equation} \label{eq:FT-time-signal}
\tilde{g}(\omega)=\mathcal{F}[g](\omega)=\sum_{\alpha}v_{\alpha}\delta(\omega-G_{\alpha})
\end{equation}
where $\delta(x)$ is the Dirac delta distribution. Hence, the uncertainty principle in Eq. (\ref{eq:standardUP}) cannot be directly applied.  

It is worth noting here that, due to the finite range of energies present in our system, there is an asymmetry between the uncertainties in time and frequency of the signal $g(t)$.  On the one hand, the uncertainty in frequency can still be well-defined. To see how, recall first that, for $g(t)  \in L^{2}$, it is possible to write the moments of a frequency signal in terms of the corresponding time signal and its derivatives, e.g.
\begin{equation}
\av{\omega^{2}} = -\frac{ \int_{-\infty}^{\infty} g^{*}(t) g''(t) dt}{ \int_{-\infty}^{\infty} |g(t)|^{2} dt}
\end{equation}
In our case, although each of these integrals diverges, their ratio does have a well-defined limit, in the sense that
\begin{equation}
\lim_{T\rightarrow \infty} -\frac{ \int_{-T}^{T} g^{*}(t) g''(t) dt}{ \int_{-T}^{T} |g(t)|^{2} dt} =  \frac{ \sum_{\alpha} |v_{\alpha}|^{2}G_{\alpha}^{2}} {\sum_{\alpha} |v_{\alpha}|^{2} } .
\end{equation}
Taking then this limit as the appropriate definition of $\av{\omega^{2}}$ in this case, and noting that, in the same sense, $\av{\omega} = 0$, we obtain that $\Delta \omega = \sqrt{\av{\omega^{2}}}$ is indeed precisely equal to  the gap dispersion $\sigma_{G}$ defined in Eq. (\ref{eq:gap-dispersion}). 

On the other hand, though, the value of $\Delta t$ diverges, even when taking limits in the same sense above. This can be understood physically due to the previously mentioned recurrences in the time signal. Indeed,  $g(t)$  is a quasi-periodic function that experiences, over an infinitely large time interval, an infinite number of recurrences to a value arbitrary close to its initial one \cite{Eberlein1949, Corduneanu2009}.

\subsection{The coarse-grained signal}

We now define the notion of \emph{coarse-graining}, in which we introduce a microscopic energy scale $\epsilon$ below which the fine-grained details of the spectrum are washed out. As we show below, this is done by replacing the discrete spectrum present in Eq. (\ref{eq:FT-time-signal}) by a suitable smooth version. 

As previously mentioned, such coarse graining of the frequency signal dampens the time signal $g(t)$, removing the recurrences seen in  Fig.~\ref{fig:time-signal} and making $g(t)$ and $\tilde{g}(\omega)$ belong to $L^{2}$. We shall see later that it will also allow us to exploit certain existing statements concerning the shapes of \emph{energy densities} and \emph{density of states} of realistic initial states
and short-ranged local Hamiltonians \cite{Brandao2015b}, which will justify our assumption of quasi-saturating the uncertainty bound.

An important issue in the coarse-graining is obviously the choice of the energy scale $\epsilon$. This will be discussed later in detail but we can already understand that in order to remove the recurrences, the discreteness of the frequency signal has to be removed, which implies an $\epsilon$ much larger than the separation between consecutive gaps.

Mathematically, the coarse-graining is accomplished by convolving the frequency signal with an appropriate \emph{window function} $h_{\epsilon}(x)$, which is only nonzero over an interval of size $O(\epsilon$). In our case, we find it convenient to choose $h_{\epsilon}(x)=C N_{\epsilon}(x)$, where 
\begin{equation}
N_{\sigma}(x)\coloneqq \frac{1}{\sqrt{2\pi}\sigma}\e^{-\frac{x^{2}}{2\sigma^{2}}}
\end{equation}
 is the normalized Gaussian distribution centred at the origin and with standard deviation $\sigma$, and
$C$ a constant that is determined below.

With this spirit, the $\epsilon$-coarse-grained
version of the frequency signal is defined as 
\begin{equation}
\tilde{g}_{\epsilon}(\omega)\coloneqq(h_{\epsilon}*\tilde{g})(\omega)=\int_{-\infty}^{\infty}\dd\omega'h_{\epsilon}(\omega'-\omega)\tilde{g}(\omega')\,.
\end{equation}

\begin{figure}
\begin{center}
\includegraphics[width=0.8\columnwidth]{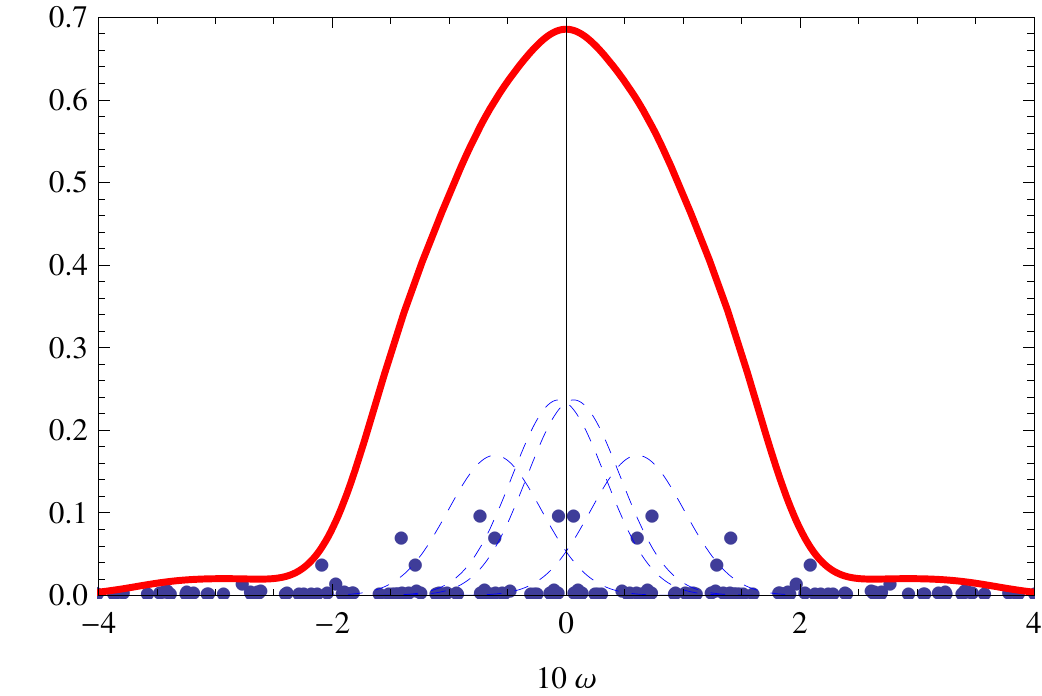} 
\caption{ (color online). Illustration of the coarse-graining of a discrete gap spectrum with a Gaussian window function. The data here corresponds to the XXZ model that is studied in section \ref{sec:XXZmodel}, with $n = 12$ spins. The solid blue dots represent the amplitudes  $v_{\alpha}(G_{\alpha})$ of each gap (in this particular case they are all real and positive). The dashed blue lines illustrate a few of the corresponding weighted Gaussians $v_{\alpha} h_{\epsilon}(\omega-G_{\alpha})$, where we have chosen $\epsilon = 0.4$ (in arbitrary frequency units). The solid red curve represents the full coarse-grained spectrum $\tilde{g}_{\epsilon}(\omega)$, obtained by summing these weighted Gaussians, according to Eq.~(\ref{eq:coarse-grained-frequency-signal}).  Note that, for this choice of $\epsilon$, the width $\Delta \omega$ of the coarse-grained spectrum remains close to the dispersion $\sigma_{G}$ of the original (discrete) gap spectrum. See Fig.~\ref{Fig-Mx-coarse-and-exact} for a comparison of the corresponding coarse-grained time signal $g_{\epsilon}(t)$ with the exact one. Note finally that although, for simplicity, we use here a single numerical scale on the vertical axis, the $v_{\alpha}$ are adimensional, whereas the continuous curves have physical dimension of time (with units that are the inverse of those used for $\omega$ and $\epsilon$). 
}
\label{fig:coarse-graining} 
\end{center}
\end{figure}

If $\tilde{g}(\omega)$ is given by Eq.~(\ref{eq:FT-time-signal}), then the $\epsilon-$coarse-grained frequency
signal is
\begin{equation}
\tilde{g}_{\epsilon}(\omega)=\sum_{\alpha}v_{\alpha}\ h_{\epsilon}(\omega-G_{\alpha})\,.\label{eq:coarse-grained-frequency-signal}
\end{equation}
In other words, coarse-graining corresponds to widening each Dirac-$\delta$  in the original spectral function into a Gaussian of $O(\epsilon)$ width (Fig~\ref{fig:coarse-graining}). Note that, in doing this, we remove  fine details of the spectrum such as the \emph{level
statistics}. Furthermore, unlike $\tilde{g}(\omega)$,   $\tilde{g}_{\epsilon}(\omega)$ is square integrable and lies in $L^{2}$ :
\begin{equation}
\int\dd\omega|\tilde{g}_{\epsilon}(\omega)|^{2}=\sum_{\alpha,\beta}v_{\alpha}v_{\beta}^{*}\ (h_{\epsilon}*h_{\epsilon})(G_{\alpha}-G_{\beta})= C^{2}\sum_{\alpha,\beta}v_{\alpha}v_{\beta}^{*}\ N_{\sqrt{2}\epsilon}\left(G_{\alpha}-G_{\beta}\right)<\infty,
\end{equation}
where we have used the fact that $(h_{\epsilon}*h_{\epsilon})(x)= C^{2} N_{\sqrt{2}\epsilon}(x)$.

A coarse-grained frequency signal defines a coarse-grained time signal
given by 
\begin{equation}
g_{\epsilon}(t)=\mathcal{F}^{-1}[\tilde{g}_{\epsilon}](t)=\mathcal{F}^{-1}[h_{\epsilon}](t)\cdot g(t)=\frac{C}{\sqrt{2\pi}}\e^{-\frac{1}{2}\epsilon^{2}t^{2}}g(t)\,,
\end{equation}
where we have used the convolution theorem for Fourier transforms. The constant $C$ is fixed
by imposing that the time signal is not affected by coarse-graining
the frequency signal in time scales $t\ll\epsilon^{-1}$ i.~e.\
$g_{\epsilon}(0)=g(0)$. This leads to $C=\sqrt{2\pi}$ and 
\begin{eqnarray}
h_{\epsilon}(\omega) &=& \sqrt{2\pi} N_{\epsilon}(\omega) =
\frac{1}{\epsilon}\e^{-\frac{\omega^{2}}{2\epsilon^{2}}}  ,\label{eq:expression-for-h}\\
g_{\epsilon}(t) &=& \e^{-\frac{1}{2}\epsilon^{2}t^{2}}g(t)\ .\label{eq:coarse-grained-time-signal}
\end{eqnarray}

Even if the time signal $g(t)$ equilibrates after some time, we know that it must eventually have recurrences. 
To determine the equilibration time-scale from $g_\epsilon(t)$, we need that $\epsilon^{-1}$ is much greater than the equilibration timescale, but much smaller than the recurrence timescale. In this way, we ensure that the coarse-grained time signal will be indistinguishable from the original one during the equilibration process, but unlike the latter will then decay to zero.

\subsection{The coarse-grained density of relevant gaps}

We now focus on the properties of the variance  of gap values with respect to the coarse-grained frequency signal $\tilde{g}_{\epsilon}(\omega)$, i.~e.\
\begin{equation}
\Delta\omega_{\epsilon}^{2}\coloneqq \frac{1}{\norm{\tilde{g}_{\epsilon}}_{2}}\int\dd\omega|\tilde{g}_{\epsilon}(\omega)|^{2}(\omega-\tilde{\mu}_{\omega})^{2}
\end{equation}
with $\tilde{\mu}_{\omega}\coloneqq \int\dd\omega|\tilde{g}_{\epsilon}(\omega)|^{2}\omega/\norm{\tilde{g}_{\epsilon}}_{2}$.
We will refer to $\Delta \omega_{\epsilon}$ as the 'dispersion of relevant gaps', and to the weight 
 \begin{equation}
|\tilde{g}_{\epsilon}(\omega)|^{2}=\sum_{\alpha,\beta}v_{\alpha}v_{\beta}^{*}h_{\epsilon}(\omega-G_{\alpha})h_{\epsilon}(\omega-G_{\beta})\label{eq:density-of-relevant-gaps}
\end{equation}
as the \emph{density of relevant gaps}.

Our goal is to show that, under a wide range of choices of $\epsilon$ and of physically relevant circumstances: i) $\Delta \omega_{\epsilon}^{2}$  is very close to the gap dispersion $\sigma_{G}^{2}$ of the original signal, as defined in Eq. (\ref{eq:gap-dispersion}), and at the same time (ii) the inverse $\Delta \omega_{\epsilon}^{-1}$ is a good estimate for the equilibration time.

Note first that, by construction, for every gap $G_{\alpha} = E_{j}- E_{i}$ in Eq. (\ref{eq:thesum}) (and, by extension, Eq. (\ref{eq:coarse-grained-frequency-signal}), its negative $G_{\bar{\alpha}} := E_{i} - E_{j} $ also appears, with  $|v_{\bar{\alpha}}| = |v_{\alpha}|$. Hence the `average gap` $\tilde{\mu}_{\omega}$ vanishes for any $\epsilon$ and the variance  $\Delta \omega ^2$ is equivalent to $\av{\omega^{2}}$, i.e.:

\begin{equation}
\Delta \omega_{\epsilon}^2 = \frac{\int \omega^{2} |\tilde{g}_{\epsilon}(\omega)|^{2} d \omega}{\int |\tilde{g}_{\epsilon}(\omega)|^{2} d \omega}
\end{equation}

After some straightforward  manipulation, using Eqs. (\ref{eq:expression-for-h}) and (\ref{eq:density-of-relevant-gaps}), we obtain
\begin{equation}
\Delta \omega_{\epsilon}^2 = \frac{\sum_{\alpha}|v_{\alpha}|^{2}\left(G_{\alpha}^{2}+\epsilon^{2}/2\right) + \frac{1}{4}\sum_{\alpha \neq \beta} v_{\alpha}v_{\beta}^{*}\left[\left(G_{\alpha} + G_{\beta}\right)^{2}+2\epsilon^{2}\right]e^{-\frac{(G_{\alpha}-G_{\beta})^{2}}{4\epsilon^{2}}} }{\sum_{\alpha}|v_{\alpha}|^{2} + \sum_{\alpha \neq \beta} v_{\alpha}v_{\beta}^{*}\; e^{-\frac{(G_{\alpha}-G_{\beta})^{2}}{4\epsilon^{2}}} }
\end{equation}

It can be easily checked that, if $\epsilon \rightarrow 0$, this expression indeed reduces to Eq. (\ref{eq:gap-dispersion}). More specifically, $\Delta \omega_{\epsilon}^2$  will be very close to $\sigma_{G}^{2}$ for all $\epsilon \ll \min(G_{\alpha} - G_{\beta})$. Indeed, in this limit the Gaussian window function $h_{\epsilon}(\omega)$ becomes negligibly thin with respect to the smallest separation between gaps, and the coarse-grained spectrum $\tilde{g}_{\epsilon}(\omega)$ resembles the original discrete spectrum $g(\omega)$. Precisely for this reason, however, this limit is of little use to our goals. Another way of putting this is that, for such small values of $\epsilon$  the coarse-grained time signal (Eq. (\ref{eq:coarse-grained-time-signal})) does not have time to decay before the recurrence timescale of the original signal, which is of order $\min(G_{\alpha} - G_{\beta})^{-1}$.

To make further progress at this point, it is necessary to assume some features
about the amplitudes $v_{\alpha}$. Otherwise, a fine tuning between
phases and modulus of $v_{\alpha}$ can make $|\tilde{g}_{\epsilon}(\omega)|$
have an arbitrary behaviour, preventing any general
statement concerning the equilibration of $g_{\epsilon}(t)$.

Inspired by \cite{Reimann2010}, we will adopt a weak-typicality point of view: let us assume 
that the evolution we are considering is drawn from an ensemble
for which the $v_{\alpha}$'s are describable by some \emph{smooth functions plus stochastic fluctuations}.
Note that we do not assume a uniform ensemble over all states (or any one specific ensemble), as is the case in most typicality studies \cite{Brandao2012,Cramer2012,Masanes2013,Malabarba2014,Goldstein2014,Goldstein2015,Reimann2016,BalzReimann2017}, merely one for which the resulting distribution over the $v_{\alpha}$'s has some very general features which are described below.
In the spirit of statistical physics, we basically \emph{replace complexity by apparent randomness}.
In most situations, the description of the gap relevances $v_\alpha$ in terms of a smooth function plus stochastic fluctuations is a consequence of the energy level populations $c_i$ and the matrix-elements of the observable $A_{ij}$ having this same behaviour.
In the next section we discuss under which conditions this is indeed the case.

In the following we show that the process of coarse-graining
removes the fluctuations and makes the density of relevant gaps $|\tilde{g}_{\epsilon}(\omega)|^{2}$
have a smooth behaviour.

\begin{result}[Coarse-grained frequency signal] 
\label{result:coarse-grained-frequency-signal}
Let us consider the
amplitudes $v_{\alpha}$ of the gaps $G_{\alpha}$ to be described by
\begin{equation}
v_{\alpha}=v(G_{\alpha})+\delta v_{\alpha}\label{eq:ansatz-v_alpha}
\end{equation}
where $v(\omega)=r(\omega)\e^{\iu\theta(\omega)}$, with $r(\omega)$
and $\theta(\omega)$ two functions with a Lipshitz constant upper
bounded by some $K \ll \epsilon^{-1}$, and where $\delta v_{\alpha}$
are independent random variables that average to zero $<\delta v_{\alpha}>=0$
and $<\delta v_{\alpha}\delta v_{\beta}>=\gamma^{2}(G_{\alpha})\delta_{\alpha\beta}$.
The variance $\gamma^{2}(\omega)$ is a function that represents the
strength of the fluctuations and also has a Lipshitz constant upper
bounded by $K$. Then, with a high
probability $\erf(m)\ge 1- \exp(-m^2)$, the density of relevant gaps fulfils the following bound 
\begin{equation}\label{eq:density-of-relevant-gaps2}
\left|\tilde{g}_{\epsilon}(\omega)-\sqrt{2\pi} v(\omega)\rho_{\epsilon}(\omega)\right|\le
\rho_{\epsilon}(\omega)\left(c_1  K \epsilon+
\pi^{1/4}m\frac{\gamma(\omega)}{\sqrt{\epsilon \rho_{\epsilon}(\omega)}}\right)\, .
\end{equation}
where $c_1>0$ is a constant, $\rho_{\epsilon}(\omega)=(\rho*N_{\epsilon})(\omega)$ is the \emph{coarse-grained
density of gaps}, with $\rho(\omega)=\sum_{\alpha}\delta(\omega-G_{\alpha})$
being the density of gaps. The coarse-grained density of gaps $\rho_{\epsilon}(\omega)$
describes how many gaps are $\epsilon$-close to the frequency $\omega$.

That is, the process
of coarse-graining washes out the fluctuations $\delta v_{\alpha}$ 
turning the coarse-grained frequency signal into a smooth function
\begin{equation} \label{eq:smooth-gap-density}
\tilde{g}_{\epsilon}(\omega)\simeq \sqrt{2\pi} v(\omega)\rho_{\epsilon}(\omega),
\end{equation}
where the meaning of the approximation is made precise in Eq.~\eqref{eq:density-of-relevant-gaps2}.
\end{result}
The proof is tedious and non-illuminating so we give it in \ref{app:proof-freq-signal}.

Note that the error in \eqref{eq:density-of-relevant-gaps2} has two components. 
The term $K\epsilon$ is due to the variation of the ``smooth'' functions with Lipshitz constant smaller than $K$
within an interval of width $O(\epsilon)$. 
The second component $\gamma (\omega)/\sqrt{\epsilon \rho_\epsilon(\omega)}$ is the fluctuation
that shrinks according to the central limit theorem with $\sqrt{\epsilon \rho_\epsilon(\omega)}$,
where $\epsilon \rho_\epsilon(\omega)$ is the number of gaps $G_\alpha$ within an interval $O(\epsilon)$.

In order for Eq.~\eqref{eq:density-of-relevant-gaps2}
to be meaningful and the error bound small, an optimization over $\epsilon$
is needed. It is easy to see that for the error to be small, the parameter $\epsilon$ should be much larger than the spacing between
consecutive gaps and much smaller than the inverse of the Lipshitz
constants of the continuous functions $v(\omega)$ and $\gamma(\omega)$,
i.~e. 
\begin{equation} \label{eq:epsilon_condition}
G_{\alpha+1}-G_{\alpha}\ll\epsilon\ll K^{-1}\,.
\end{equation}
Here `consecutive' refers to gaps ordered by size, and we use the informal notation ``$G_{\alpha +1}$''  as a shorthand for ``the gap immediately larger than $G_\alpha$''.  Note that, in practice, the lower bound above should be applied only for consecutive \emph{relevant} gaps, i.e., gaps that whose amplitudes $v_{\alpha}$ make a non-negligible contribution to the sum in Eq. (\ref{eq:thesum}). 

For many-body systems, the number of gaps increases exponentially in the system size $n$, and energy differences between consecutive gaps shrink exponentially
to zero in $n$. If for example these gaps all have roughly equal (exponentially small) relevances $q_{\alpha}$, then
$\epsilon$ can also be taken exponentially to zero, as long as this is done at a slower rate than the difference in gaps. This makes the time signal and its coarse-grained version indistinguishable 
in any realistic time-scale.

In sum, 
if the gap relevances $v_\alpha$ 
can be described by a continuous part plus a fluctuating part, as in Eq.~\eqref{eq:ansatz-v_alpha}, then the coarse-grained density of relevant gaps is a smooth function given by
\begin{equation} 
|\tilde{g}_{\epsilon}(\omega)|^2\simeq 2\pi |v(\omega)|^2\rho_{\epsilon}(\omega)^2\, . 
\end{equation}
This will be particularly useful in the following section, where we apply these ideas in 
the case of many body systems described by short-ranged Hamiltonians.

It is worth noting that the factorization in Eq.(\ref{eq:smooth-gap-density}) is also automatically obtained if one assumes, as is often done, that in the thermodynamic limit $n \rightarrow \infty$  the discrete gap spectrum may be replaced by a smooth continuous gap density. i.e. taking
\begin{align}
g(t) = \sum_{\alpha} v_{\alpha}e^{\iu G_{\alpha}t} \rightarrow \int  v(\omega) e^{\iu \omega t} \rho_{G}(\omega) d \omega
\end{align}
Comparing with the definition in Eq. (\ref{eq:freq-signal}), we see that in this case the frequency signal is again $\tilde{g}(\omega)~=~\sqrt{2\pi}~v(\omega)  \rho_{G}(\omega) $. This indicates that the assumptions we have made concerning the smoothness of $v_{\alpha}$ are not severe. However, the point of attaining this relation via coarse-graining, while maintaining a finite dimension $n$, is that it allows us to control how the equilibration timescale scales with the system size, and in particular to understand how this scaling depends on the energy scaling of the relevant observable. We turn to this question in the next section.

\section{Relevant energy scales and equilibration time scales for local Hamiltonians}
\label{sec:energyscales}

In this section, we focus on the particular but relevant case of short-range
Hamiltonians and initial states that have a finite correlation length.
For such systems, we express the density of relevant gaps in terms
of the energy density of the initial state and the function that describes
the matrix-elements of the observable. 
By doing so, we identify the energy scales
that determine the dispersion of gaps. We find that there are mainly
two relevant energy scales: the energy fluctuations of the initial
state and the bandwidth of the matrix of the observable $A$ in the
Hamiltonian eigenbasis. In the case of systems globally out of equilibrium,
only those observables that are banded in the Hamiltonian basis can
be observed out of equilibrium for a non-negligible time.

\vspace{.5cm}
\noindent
\textbf{Local Hamiltonian}. Let us define a \emph{short-ranged} or \emph{local} Hamiltonian of
a spin lattice system, i.e., acting on a Hilbert space $\hiH=\bigotimes_{x\in V}\hiH_{x}$
with $\dim(\hiH_{x})=d$, as 
\begin{equation}
H=\sum_{u\in\mathcal{E}}h_{u}\label{eq:local-hamiltonian}
\end{equation}
where the locality structure is given by a graph $(V,\mathcal{E})$
with a vertex set $V$ and edge set $\mathcal{E}$. The number of
terms of the Hamiltonian is denoted by $n=|\mathcal{E}|$. We consider
systems for which it is possible to define a sequence of Hamiltonians
$H_{n}$ of different sizes. This becomes trivial in the case of translational
invariant systems and regular lattices, but also includes systems
with disorder, and defects. The reason for introducing such a sequence
of Hamiltonians $H_{n}$ is that it allows us to define the thermodynamic
limit. For simplicity, the subindex $n$ is not explicitly written
from now on.

\vspace{.5cm}
\noindent
\textbf{Energy density of the initial state}.
The \emph{energy density} $f(E)$ of an initial state $\ket{\psi(0)}=\sum_{i}c_{i}\ket{E_{i}}$
is defined by 
\begin{equation}
f(E):=\sum_{i=1}^{d_{E}}|c_{i}|^{2}\delta(E-E_{i})\,,
\end{equation}
and, what is more relevant for us, its coarse-grained version reads
$f_{\epsilon}(E):=(N_{\epsilon}*f)(E)$.

For the case of local Hamiltonians, the coarse-grained energy density
of states with a finite correlation length has been proven to approach a Gaussian  as the size $n$ of the system increases
(Lemma 8 in Ref.~\cite{Brandao2015b}). More specifically, 
 for all $E$,
\begin{equation}
\int_{-\infty}^E dE' f_{\epsilon}(E') \rightarrow
\int_{-\infty}^E dE' \frac{1}{\sqrt{2\pi}\sigma_{E}}\e^{-\frac{1}{2}\frac{(E'-\mu_{E})^{2}}{2\sigma_{E}^{2}}}
\label{eq:Gaussian-energy-density}
\end{equation}
where the Gaussian has  mean $\mu_{E}=\bra{\psi(0)}H\ket{\psi(0)}$ and standard deviation $\sigma_{E}^{2}~=~\bra{\psi(0)}(H-~\mu_{E})^{2}\ket{\psi(0)}$, and where the difference in these expressions falls, at worst, essentially as $1/\sqrt{n}$. In what follows, we only make use of the energy density inside integrals, 
so in practice Eq.~\eqref{eq:Gaussian-energy-density} allows us to replace the energy density $f_{\epsilon}(E')$ by the corresponding Gaussian, with vanishing error.

Initial states that
are globally out of equilibrium, e.~g.\ globally quenched, have
energy densities with mean and standard deviation that scale in the
system size as $\sigma_{E}\propto\sqrt{n}$ and $\mu_{E}\propto n$.

If the system is at criticality and the correlations decay in a power
law, this Gaussian shape cannot be guaranteed anymore. In any case, the
energy fluctuations can still scale as $\sigma_{E}\propto\sqrt{n}$
as long as the power $m$ of the decay is sufficiently fast, i.~e.\ $m>D+1$
where $D$ is the spatial dimension of the lattice (see \ref{sec:local-Ham}
for details).

\vspace{.5cm}
\noindent
\textbf{Matrix-elements of an observable in the energy basis}.
Taking again a weakly stochastic approach, in the same spirit of the argument used in Result 2, we constrain ourselves to observables
with off-diagonal matrix-elements in the Hamiltonian basis that can
be described by a ``continuous'' function $S(E,\omega)$ plus some
fluctuations $\delta A_{ij}$ 
\begin{equation}
|A_{ij}|^{2}=S\left(\frac{E_{i}+E_{j}}{2},\frac{E_{i}-E_{j}}{2}\right)+\delta A_{ij}\,,
\label{eq:definition-of-S}
\end{equation}
where this choice of writing the arguments of the function $S(E,\omega)$
will be shown to be convenient in the following section. 
As in the previous section, the
Lipshitz constant of $S(E,\omega)$ is assumed to be bounded by $K$.

Note that the so called Eigenstate Thermalization Hypothesis (ETH) 
\cite{Srednicki1994} can be seen as a
particular case of this assumption \eqref{eq:definition-of-S}.
One popular version of the ETH \cite{Dalessio2016} is an ansatz on the matrix-elements
of an observable $A$ in the Hamiltonian eigenbasis, 
\begin{equation}
A_{ij}=\bra{E_{i}}A\ket{E_{j}}=\mathcal{A}(E)\delta_{ij}+\rho(E)^{-1}f_{\textrm{ETH}}(E,\omega)R_{ij}\label{eq:ETH}
\end{equation}
where $E=(E_{i}+E_{j})/2$ and $\omega=(E_{i}-E_{j})/2$. The functions
$\mathcal{A}$ and $f_{\textrm{ETH}}(E,\omega)$ are smooth functions of their arguments,
and $R_{ij}$ are complex numbers randomly distributed, each with
zero mean and unit variance.
The essential idea is that both in \eqref{eq:definition-of-S} and \eqref{eq:ETH}
the off-diagonal matrix-elements of the observable can be described by a smooth function plus fluctuations that vanish when coarse-graining. 

Note that in our case, in contrast to ETH, we do not assume anything about the diagonal elements of the
observable in the energy basis. This is due to the fact that we are not concerned about what is the equilibrium state of the system (whether is thermal or not), but only about how long the relaxation process takes.

Now that we have introduced the energy density $f_{\epsilon}(E)$ and the function $S(E,\omega)$
that describes the observable, we are ready to express the density of relevant gaps in terms of these functions:

\begin{result}[Density of relevant gaps] Given a Hamiltonian and
an initial state with populations $|c_{i}|^{2}$, let A be
an observable  whose matrix-elements in the Hamiltonian
eigenbasis, $|A_{ij}|^{2}$ , can be described by means
of the smooth function $S(E,\omega)$ plus some fluctuations as in Eq.~\eqref{eq:definition-of-S}. 
Then, up to errors $O(\epsilon K)$, the
density of relevant gaps can be written as 
\begin{equation}
|\tilde{g}_{\epsilon}(\omega)|^{2}=2\pi\rho_{\epsilon}(\omega)\ \int_{-\infty}^{\infty}\dd Ef_{\epsilon}(E-\omega/2)f_{\epsilon}(E+\omega/2)S(E,\omega)\,.\label{eq:relevant-gaps-vs-other-densities}
\end{equation}
Furthermore, if the Hamiltonian is local and the initial state has
a finite correlation length, such that the coarse-grained energy density
$f_{\epsilon}(E)$ is the Gaussian \eqref{eq:Gaussian-energy-density},
then 
\begin{equation}\label{eq:local-density-relevant-gaps}
|\tilde{g}_{\epsilon}(\omega)|^{2}=2\pi \, N_{\sqrt{2}\sigma_{E}}(\omega)\,  S(\omega)\rho_{\epsilon}(\omega)\, ,
\end{equation}
where the function $S(\omega)$ is defined as 
\begin{equation}
S(\omega)\coloneqq \int\dd E\, N_{\frac{\sigma_{E}}{\sqrt{2}}}(E-\mu_{E})S(E,\omega)\, ,
\end{equation}
with $\mu_{E}$ and $\sigma_{E}$ the mean and standard deviation
of the energy density. 
\label{Result3}
\end{result} 
\begin{proof} Let us introduce
the density 
\begin{equation}
R_{\epsilon}(\omega)\coloneqq \int_{-\infty}^{\infty}\dd E\, f_{\epsilon}(E-\omega/2)f_{\epsilon}(E+\omega/2)S(E,\omega)\, \label{eq:a-useful-density}
\end{equation}
and plug in it the energy density $f_{\epsilon}(E):=\sum_{i=1}^{d_{E}}|c_{i}|^{2}N_{\epsilon}(E-E_{i})$.
A straightforward calculation leads to 
\begin{equation}
R_{\epsilon}(\omega)=\sum_{i,j}N_{\sqrt{2}\epsilon}\left(\omega-(E_{i}-E_{j})\right)|c_{i}|^{2}|c_{j}|^{2}\int\dd E\, S(E,\omega)N_{\frac{\epsilon}{\sqrt{2}}}\left(E-\frac{E_{i}+E_{j}}{2}\right)\,,
\end{equation}
where we have used that $N_{\epsilon}(E+\omega-E_{i})N_{\epsilon}(E-\omega-E_{j})=N_{\frac{\epsilon}{\sqrt{2}}}\left(E-\frac{E_{i}+E_{j}}{2}\right)N_{\sqrt{2}\epsilon}\left(\omega-(E_{i}-E_{j})\right)$.
By considering that $S(E,\omega)$ is the smooth description of the
matrix elements $|A_{ij}|^{2}$, we get, up to errors of order $\epsilon K$,
\begin{equation}\label{eq:ref-R}
R_{\epsilon}(\omega)=\sum_{\alpha}|v_{\alpha}|^{2}N_{\sqrt{2}\epsilon}(\omega-G_{\alpha})=|v(\omega)|^{2}\rho_{\sqrt{2}\epsilon}(\omega)\, .
\end{equation}
Putting Eqs.~\eqref{eq:a-useful-density} and \eqref{eq:ref-R} together 
with Result 3 implies \eqref{eq:relevant-gaps-vs-other-densities}.

Now for initial states with a Gaussian energy density as in Eq.~\eqref{eq:Gaussian-energy-density},
simple algebra leads to the identity 
\begin{equation}
f_{\epsilon}(E-\omega/2)f_{\epsilon}(E+\omega/2)=N_{\frac{\sigma_{E}}{\sqrt{2}}}(E-\mu_{E})N_{\sqrt{2}\sigma_{E}}(\omega)\,.\label{eq:identity-gaussians}
\end{equation}
Plugging \eqref{eq:identity-gaussians} in \eqref{eq:relevant-gaps-vs-other-densities}
completes the proof. \end{proof}

The function $S(\omega)$ describes the average magnitude
of the off diagonal matrix-elements $|A_{ij}|^{2}$ at a distance
$\omega=E_{i}-E_{j}$ from the diagonal.

Result \ref{Result3} and in particular Eq.~\eqref{eq:local-density-relevant-gaps}
show that the density of relevant gaps $|\tilde{g}_\epsilon(\omega)|^2$
for local Hamiltonians and initial states with decaying correlations
decomposes in the product of two densities: $S(\omega)\rho_{\epsilon}(\omega)$ 
and $N_{\sqrt{2}\sigma_{E}}(\omega)$.
The density $N_{\sqrt{2}\sigma_{E}}(\omega)$ is a Gaussian with standard
deviation controlled by the energy fluctuations of the initial state
$\sigma_{E}$, and $S(\omega)\rho_{\epsilon}(\omega)$ is
the density of the off diagonal elements of the observable $A$.

The dispersion of relevant gaps $\sigma_{G}$, which we expect to estimate the
equilibration time, is then controlled by the
smallest of the standard deviations of these two densities. 
In the case that the system is globally out of equilibrium, e.~g., a global
quench, the variance of the energy density of the initial state is
extensive with the system size, and $\sigma_{E}\propto\sqrt{n}$.
This implies the following statement:

\begin{result}[Out of equilibrium observables in global quenches]
Given a local Hamiltonian, and an initial state with clustering of correlations
let $A$ be an observable that can be described by a smooth function
$S(E,\omega)$. Then, the only one way to avoid that 
the dispersion of relevant gaps $\sigma_G$ associated to
an observable $A$ diverges in the macroscopic limit is that
the matrix-representation of the observable $A$ in the energy basis is banded.
More specifically,
the density $S(\omega)\rho_\epsilon(\omega)$
 has a standard deviation $\sigma_{A}$ that is independent
of the system size 
\begin{equation}
\lim_{n\to\infty}\sigma_{A}=\lim_{n\to\infty}\frac{\int_{\epsilon}^{\infty}\dd\omega\,(\omega-\mu_{A})^{2}S(\omega)\rho_{\epsilon}(\omega)}{\int_{\epsilon}^{\infty}\dd\omega\,S(\omega)\rho_{\epsilon}(\omega)}<\infty\, , \label{eq:physically-relevant-condition}
\end{equation}
where $\mu_{A}$ is the first moment. 
\end{result}

Note that a divergent dispersion of relevant gaps $\sigma_G$ is
expected to imply an equilibration time that tends to zero in the thermodynamic limit.
In other words, observables which are not banded in their energy representation
are expected to be always equilibrated, 
since the amount of time that they can be out of equilibrium is negligible.

It is worth mentioning that generic observables have a flat $S(\omega)$ and
fulfil $\sigma_{A}\propto\sqrt{n}$
due to the domination of the density of gaps $\rho_{\epsilon}(\omega)$.
Thus, they will have microscopically short equilibration times \cite{Tasaki2013}.
Observables with the property \eqref{eq:physically-relevant-condition}
turn out to be both rare and physically relevant. 

In Ref. \cite{Beugeling15} it is shown for several concrete
examples that indeed the matrix elements $S(\omega)$
decrease exponentially or super-exponentially with $\omega$ from
a certain threshold independent of the system size (see also Sec. 4.3.1.2 of \cite{Dalessio2016}). 
In the following result, we show that this is a property of 
any local observable, that is, 
operators that only act non-trivially on a finite region of the
system.

\begin{result}[Local operators are banded in the energy basis]
\label{result:local-operators}
Let us consider a local Hamiltonian 
\begin{equation}
H=\sum_{(x,x')\in\mathcal{E}}h_{(x,x')}
\end{equation}
acting on a Hilbert space $\hiH=\bigotimes_{x\in V}\hiH_{x}$ with
$\dim(\hiH_{x})=d$, with a locality structure given by a graph
with a vertex set $V$ and edge set $\mathcal{E}$. Then, the matrix
elements in the energy eigenbasis of a local operator $\bra{E_{i}}A_{x}\ket{E_{j}}$
acting on a site $x$ fulfil the condition 
\begin{equation}
|\bra{E_{i}}A_{x}\ket{E_{j}}|\le \norm{A_x}\e^{\log\left(\frac{\e(E_i-E_j)}{J(1+\alpha)}\right) -c (E_{i}-E_{j})/J}
\end{equation}
where $c=\log(1+\alpha^{-1})$ is the decay rate, $J=\max_{(x,x')\in\mathcal{E}}\norm{h}_{\infty}$ is the strength
of the local interactions and $\alpha$ is the \textbf{lattice animal constant} \cite{Miranda2011} of the graph
 $(V,\mathcal{E})$.
 \end{result} 
 
The proof of Result \ref{result:local-operators} is presented in \ref{app:proof-local-operators}. In particular, note that
the lattice animal constant  mentioned above is a parameter that captures the connectivity of the underlying graph of the Hamiltonian. 
For $D$-dimensional cubic lattices, it can be bounded as $\alpha \leq 2\,D\,\e$ (Lemma~2 in Ref.~\cite{Miranda2011}).

In the period of finishing this manuscript we have been alerted to the existence of a result very similar to our Result \ref{result:local-operators}, due  to Arad et al. (Theorem 2.1 in Ref.~\cite{Arad2016}).
Both proofs are similar in spirit and 
give similar decay rates. For a $D$-dimensional cubic lattice with interactions in the edges,
they obtain a decay rate of $1/(8D)$
while we get $\log(1+(2\e D)^{-1})\simeq 1/(2 \e D)$.
The main difference is that while we bound the number of terms by counting lattice animals,
they use a combinatoric argument. 

Of course, this behaviour of being banded in the energy basis extends to global operators that can be decomposed into a sum of local terms, as well as for operators that are not local in real space but in momentum space when the Hamiltonian is also local in the momentum representation.
Note that, indeed, most observables considered in the literature are of this type. 

Let us now consider the relevant scenario of a \emph{local quench} \cite{Polkovnikov11},
in which the system is brought out of equilibrium in only a local region
of the system. In such a case, the width of the energy density of
the initial state is independent of the system size and related to
the operator norm of the perturbation applied on the system. Unlike
the global quench scenario, now even the observables that are not banded (and are initially out of equilibrium) will take a finite non-negligible time to relax.
The equilibration timescale is then governed by whichever energy scale is smallest: the energy fluctuations of the state, or the dispersion $\sigma_{A}$ of the observable.
Note that our results also allow for having equilibration times that increase with the system size, as long as either $\sigma_A$ and $\sigma_E$ shrink with it. 

\section{Numerical example: the XXZ model} \label{sec:XXZmodel}

\begin{figure}[h]
\centering \includegraphics[scale=1.0]{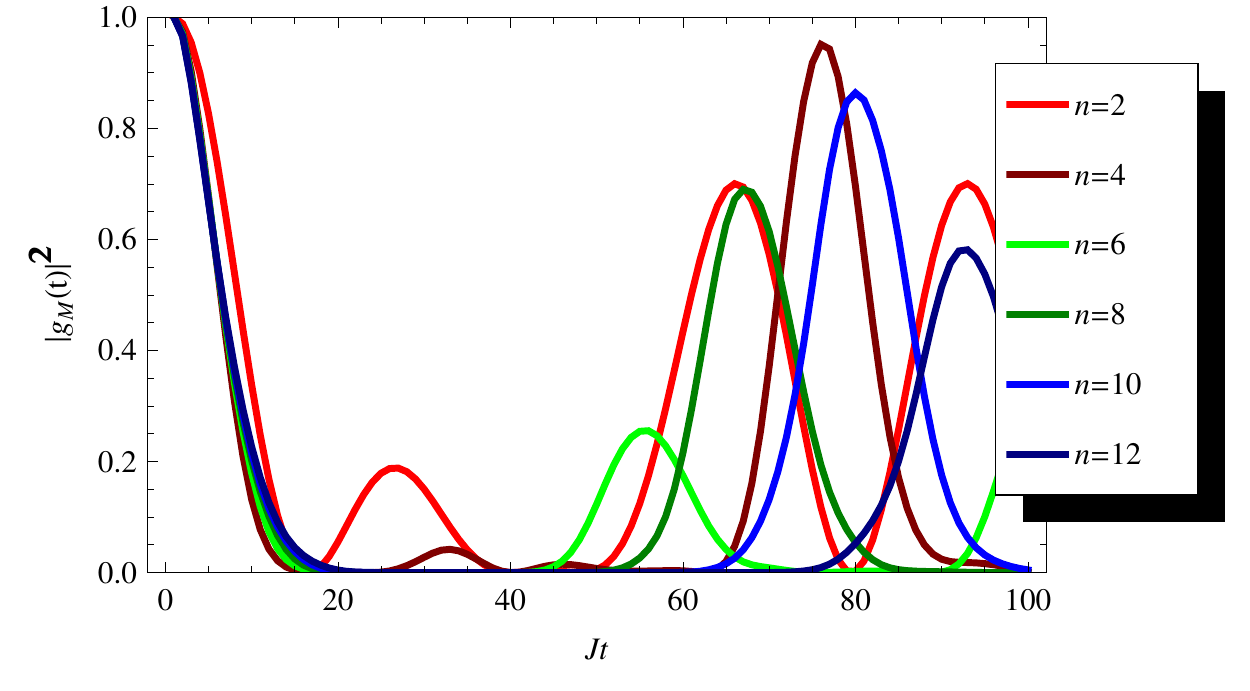}
\caption{(Color online) Fluctuation in the total magnetization in the
$x$ direction, $|g_M(t)|^{2}$, for the XXZ model with next-nearest-neighbour coupling and external magnetic field ($\Delta=0.5, J_2=1.0$ and $h_z=0.2$). The different curves are for chains with different numbers $n$ of spins.}
\label{Fig-Mx}
\end{figure}

We illustrate our results using the XXZ model in a transverse field and with next-nearest-neighbour coupling. We choose to use this particular model since it is not integrable, and hence does not have an exponential number of degenerate gaps. The hamiltonian is $H=J \sum_{i}\,\, S_{i}^{x}S_{i+1}^{x}+S_{i}^{y}S_{i+1}^{y}+\Delta S_{i}^{z}S_{i+1}^{z}+J_2 S_{i}^{z}S_{i+2}^{z}+h_z S_{i}^{z}$.
As our equilibrating observable we choose the magnetization density in the $x$ direction, $M^{x}=\sum_{i}S_{i}^{x}/N$ and, as our initial state, the fully magnetized state in the $x$ direction. 
In Fig.~\ref{Fig-Mx}, we show the evolution of the time signal $|g_M(t)|^{2}$ of this observable, in the sense of Eq. (\ref{eq:timesignal}). The calculations were done using full exact diagonalization of $H$
with $\Delta=0.5, J_2=1.0$ and $h_z=0.2$, for various
system sizes. 

We expect $g_M(t)$ to go to zero when the system equilibrates. Indeed, this is what happens initially, for all system sizes, and we can notice that the equilibration time (the time when $|g_M(t)|^{2}$ becomes negligible, just before $Jt = 20$) does not depend much on $n$.  Furthermore, we can compare this value with our heuristic estimate for the equilibration time, $ \teq \sim \pi/\sigma_G$, where we use  Eq.~(\ref{eq:gap-dispersion}) to calculate $\sigma_G$ from the numerically obtained eigenvalues.
The results are shown in table \ref{Tab-EqTimes}. We can see that the estimated equilibration
times also depend only weakly on $n$, and are in good agreement with the timescale indicated by Fig.~\ref{Fig-Mx}.

\begin{table}[h]
\centering %
\begin{tabular}{|c|c|c|c|c|}
\hline 
$n$  & $\teq \sim \pi/ \sigma_{G}$ \tabularnewline
\hline 
\hline 
2   & 21   \tabularnewline
\hline 
4   & 19   \tabularnewline
\hline 
6   & 20   \tabularnewline
\hline 
8   & 22   \tabularnewline
\hline 
10  & 23   \tabularnewline
\hline 
12  & 24   \tabularnewline
\hline 
\end{tabular}
\caption{\label{Tab-EqTimes} Estimated equilibration times $\teq$ for the XXZ model with next-nearest-neighbour coupling and an external magnetic field ($\Delta=0.5, J_2=1.0$ and $h_z=0.2$). The gap dispersion $\sigma_{G}$ was obtained by explicitly calculating Eq.~\eqref{eq:gap-dispersion} from the numerically obtained energy spectrum.}
\end{table}

Of course, due to the small size of the simulated systems, the time signals $g_M(t)$ also exhibit strong fluctuations. However, one can already see that,  as $n$ increases, the size of these fluctuations tends to decrease, and their onset happens later. Our numerical results therefore seem to corroborate our expectation that the observable $M^{x}$ does indeed equilibrate in the limit of large $n$, and that this equilibration does happen at a timescale roughly given by  $ \teq \sim \pi/\sigma_G$.

\begin{figure}[h]
\centering 
\includegraphics[scale=0.4]{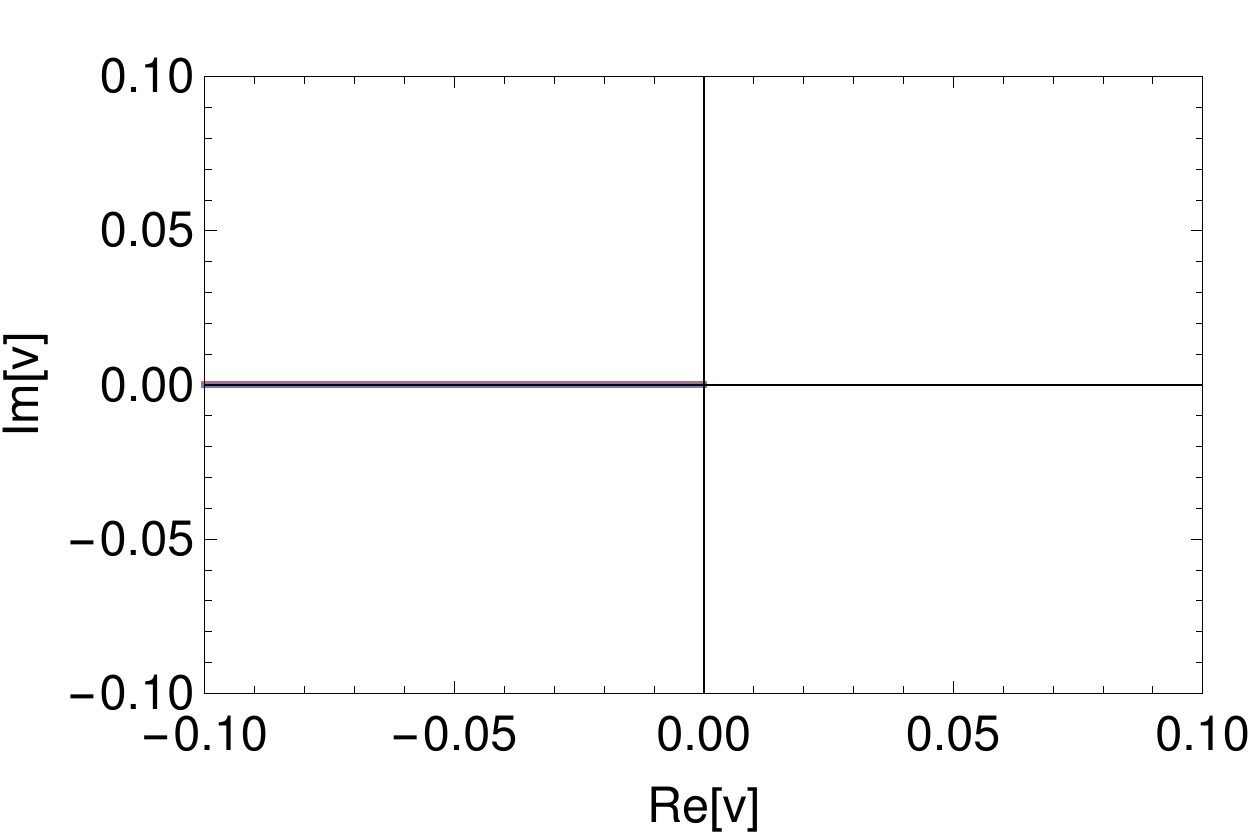}
\includegraphics[scale=0.4]{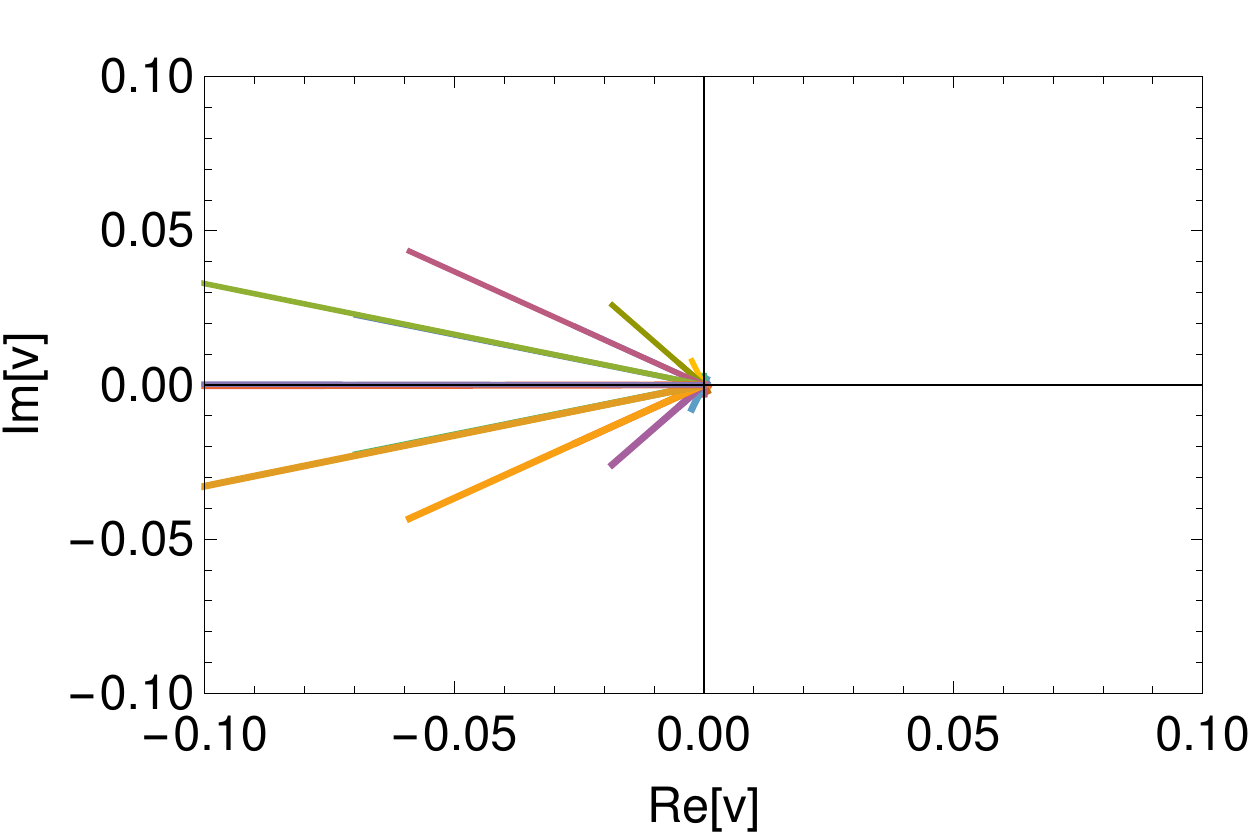}
\includegraphics[scale=0.4]{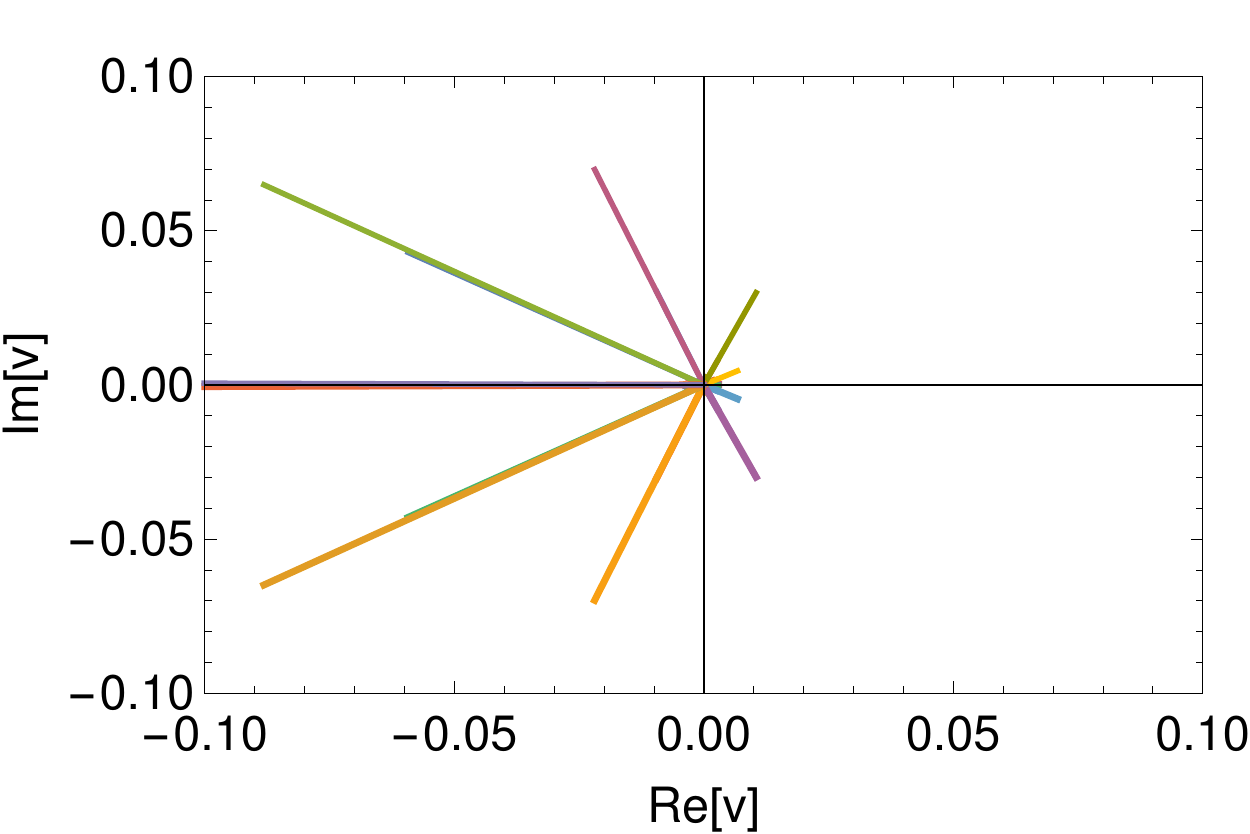}
\\
\includegraphics[scale=0.4]{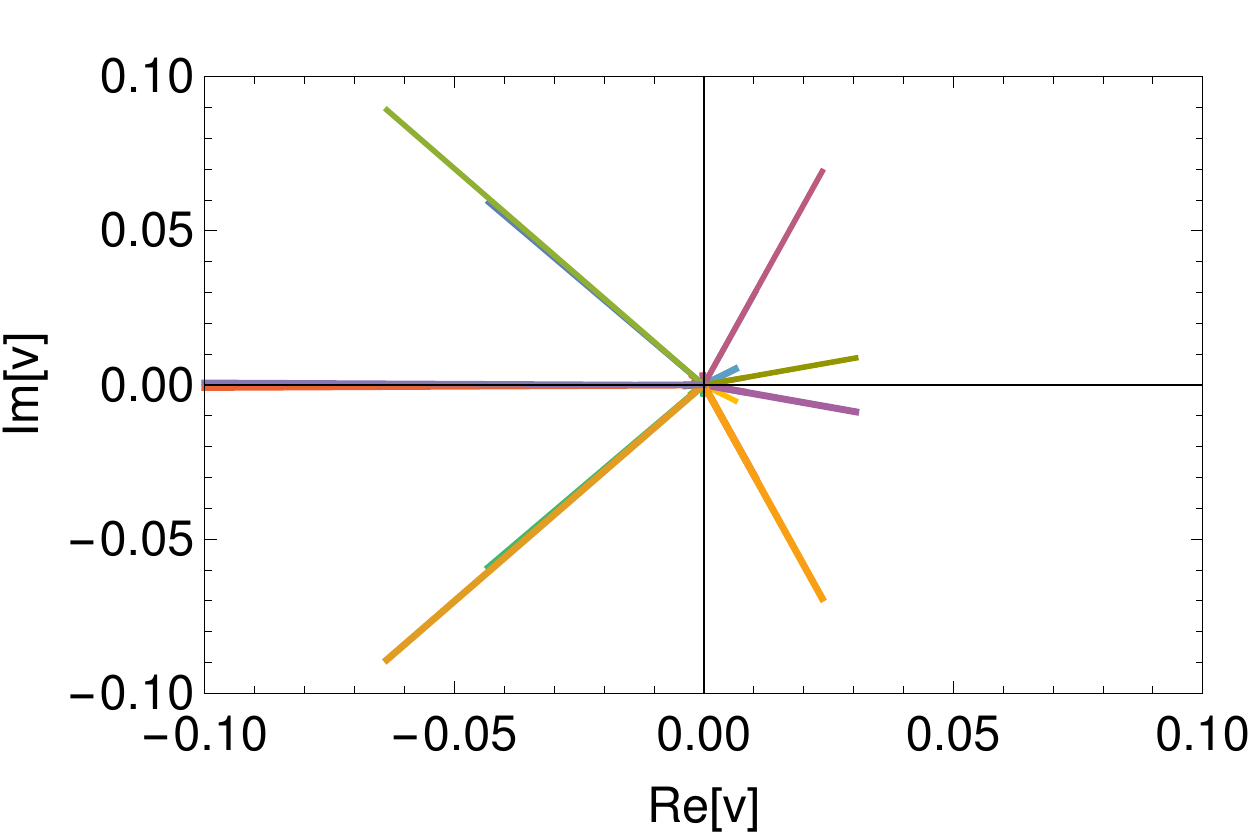}
\includegraphics[scale=0.4]{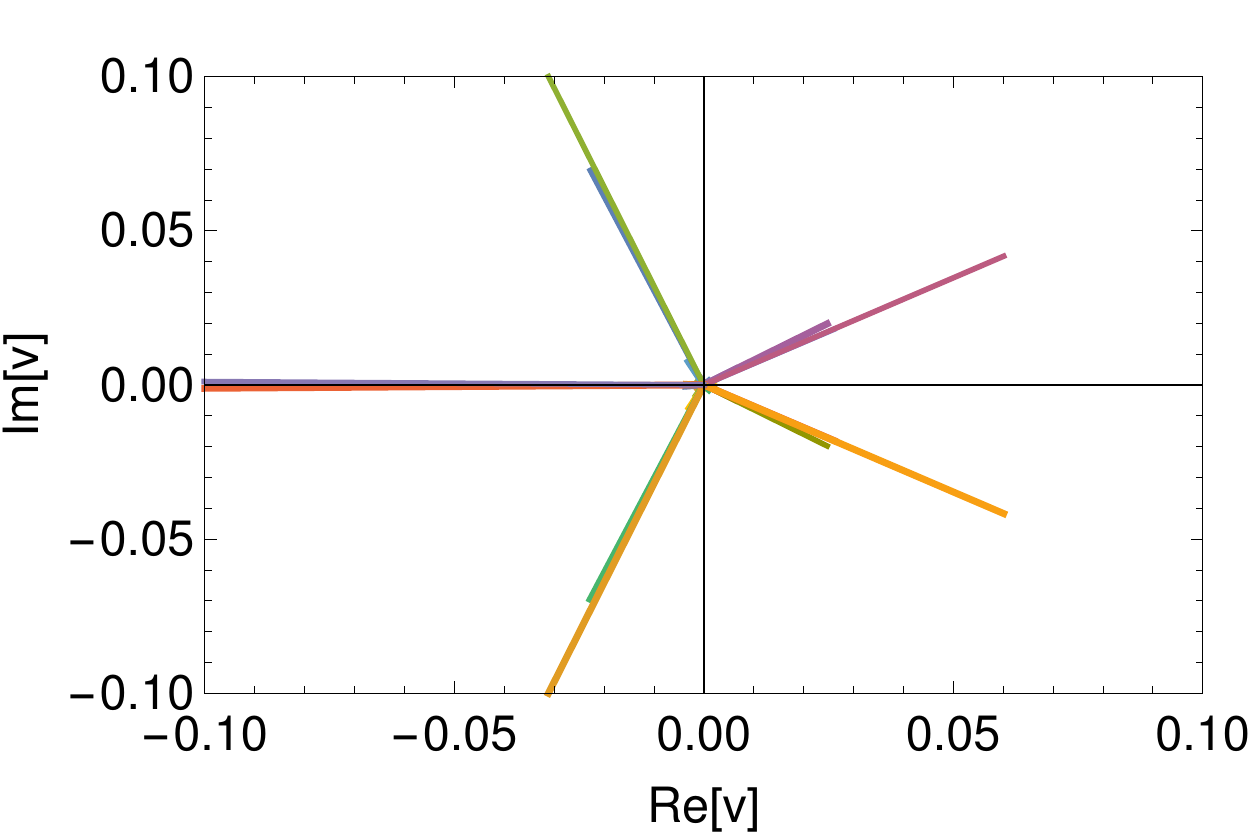}
\includegraphics[scale=0.4]{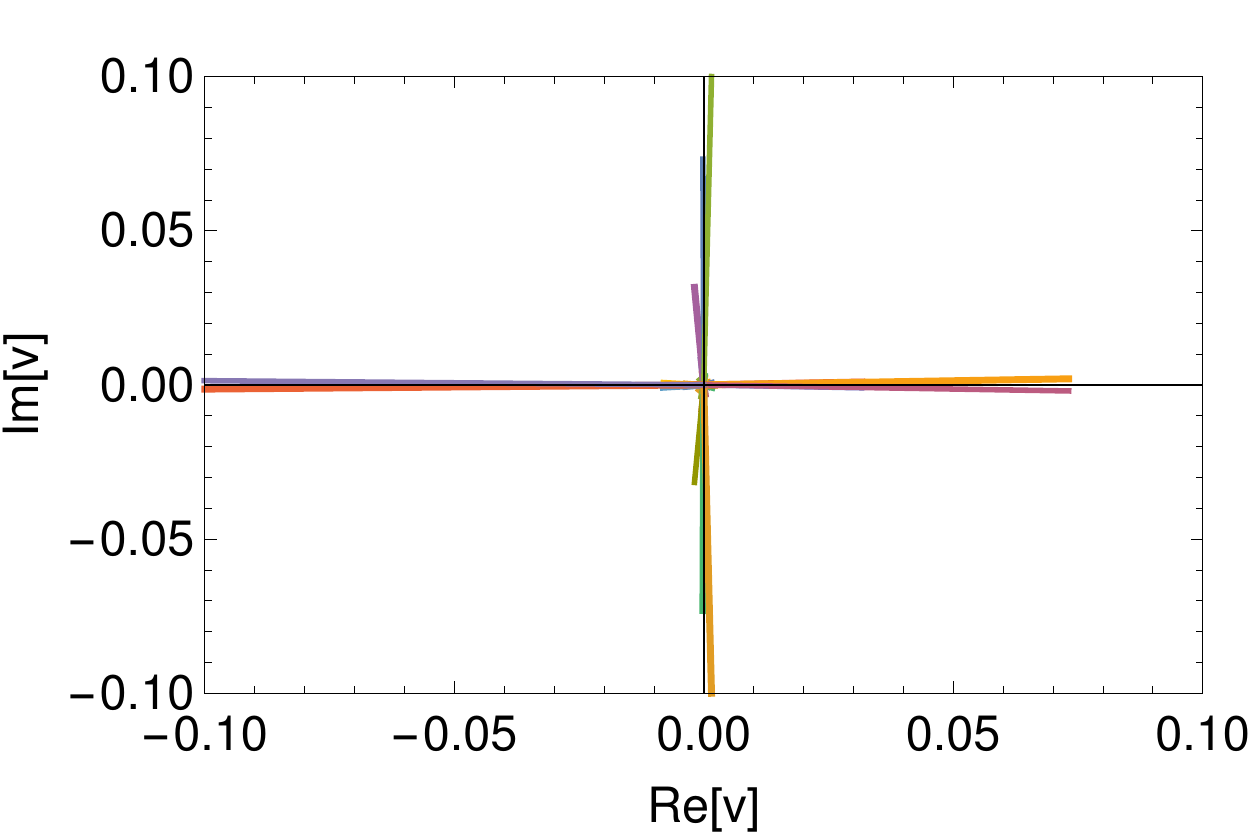}
\caption{(Color online) Evolution of each complex number $v_\alpha \e^{\iu G_{\alpha}t}$ for our simulation of the XXZ model with $n=10$. From upper left to the bottom right, (adimensional) time $Jt$ goes from 0 to 20 in equal intervals.}
\label{Fig-Deph-XYZ}
\end{figure}

To better illustrate the dephasing mechanism behind the equilibration process, in Fig.~\ref{Fig-Deph-XYZ} we plot the amplitudes $v_{\alpha}\e^{\iu G_{\alpha}t}$ for this same situation, in the case $n=10$. Starting from an initial condition where all the amplitudes are in phase (in this case, all real and negative), one can see them rotating at different speeds and becoming more spread out in the complex plane as time goes by, resulting in the decay seen in Fig.~\ref{Fig-Mx}. Note that the approximate four-way symmetry exhibited at $Jt=20$ (implying $g(t) \simeq 0$ at this time) is already a symptom of a future recurrence: clearly, after four times this interval, all of the amplitudes will have rotated approximately back to their initial position. Indeed, one can see in Fig.~\ref{Fig-Mx} that a recurrence occurs at around $Jt = 80$.

\begin{figure}[h]
\centering \includegraphics[scale=1.0]{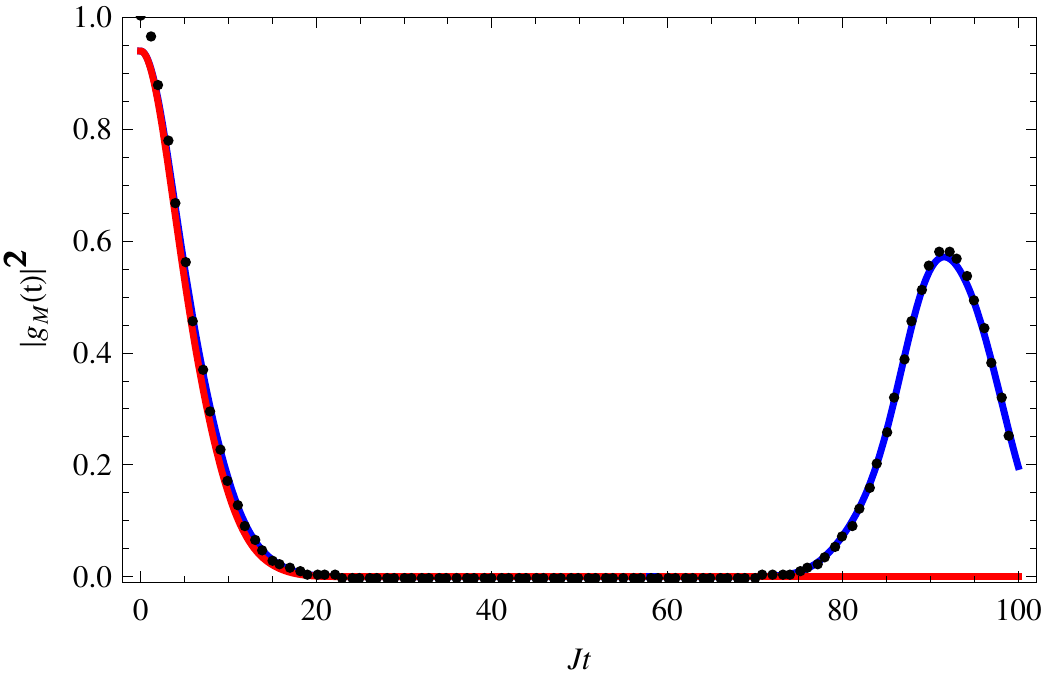}
\caption{(Color online) Comparison of exact and coarse-grained  time signals for the magnetization. The exact time signal  $g_M(t)$ (black dots) is the same curve plotted in Fig.~\ref{Fig-Mx}, for $n = 12$ spins. The full curves are coarse-grained time signals $g^{\epsilon}_M(t)$, obtained by Fourier transforming the coarse-grained frequency signals calculated according to Eq.~(\ref{eq:coarse-grained-frequency-signal}), with $\epsilon = 0.4$ (red) and  $\epsilon = 0.02$ (blue). By choosing a value of $\epsilon$ that is sufficiently large (but not too large), the coarse-grained signal reproduces the exact one during the equilibration phase, but suppresses later recurrences.}
\label{Fig-Mx-coarse-and-exact}
\end{figure}

It is also instructive to compare the exact time evolution of the magnetization with coarse-grained versions derived according to the procedures described in section \ref{sec:Fourier}. In Fig.~\ref{Fig-Mx-coarse-and-exact} we plot again the exact time signal $g_M(t)$ for the chain with $n=12$ spins (the dark blue line in Fig.~\ref{Fig-Mx}), represented here by the black dotted curve.  We also plot two coarse-grained time signals $g^{\epsilon}_M(t)$, with $\epsilon = 0.4$ (red line) and  $\epsilon = 0.02$ (blue line). These curves were obtained by Fourier transforming coarse-grained frequency spectra (such as the one in Fig.~\ref{fig:coarse-graining}, that were numerically calculated according to Eq.~(\ref{eq:coarse-grained-frequency-signal})
(i.e., we did not simply dampen the exact time signal using Eq.~(\ref{eq:coarse-grained-time-signal})). It can be seen that all three signals are essentially indistinguishable up to the equilibration time (the slight deviation close to $t = 0$ is an artifact of our having discarded terms with $v_{\alpha} < 10^{-4}$ when calculating the sum in Eq.~(\ref{eq:coarse-grained-frequency-signal})). We can see that, for small $\epsilon$ $(=0.02)$, the signals remain indistinguishable up to and including the recurrence time. However, by choosing a value of $\epsilon$ that is sufficiently large (0.4), the coarse-grained signal faithfully reproduces the exact one during the equilibration phase, but suppresses later recurrences.  Note that $\epsilon$ must still be chosen sufficiently small ($\epsilon \ll \sigma_{G}$) in order to avoid suppressing the signal even before equilibration has occurred.  As discussed in previous sections, under these circumstances we may use the width of the (square-integrable) coarse-grained signal as a measure of the equilibration time of the original signal.

Finally, let us remark that we also obtained similar results for
other values of $\Delta$ and $h_z$, and also for the XY model. However, in the latter case, the  fluctuations do not decrease exponentially with $n$, but only polynomially, since there are an exponential number of degenerate gaps.

\section{Discussion}\label{sec:discussion}
\subsection{Reinterpretation of previous results}
\label{subsec:reinterp}
It is useful now to reinterpret some previous results on equilibration times from our dephasing point of view.
For example, Short and Farrelly \cite{Short2012} obtain a rigorous upper bound for the equilibration time of any observable by studying the time-averaged fluctuations $\av{|g|^{2}}_{T} $ in Eq. \eqref{eq:finite-time-average-distance} above. They are able to determine a value $T_{0} \gtrsim d_{E}/ \Delta E$, where $\Delta E $ is the range of energies in the system, and $d_{E}$ the number of different energy levels, such that averages taken over intervals longer than $T_{0}$ are negligible. This implies that the equilibration time must be upper bounded by $T_{0}$. Unfortunately, for a typical many-body system with $n$ degrees of freedom, $\Delta E$ scales only polynomially with $n$, while $d_{E}$ scales exponentially with $n$ - and thus so does $T_{0}$. In other words, although this upper bound is mathematically sound, it vastly overestimates the actual equilibration time of most observables. This suggests that its derivation must be incomplete, in the sense of missing or disregarding an essential physical ingredient \cite{Tasaki2013}. 

We now argue that this ingredient is, in a word, dephasing. Roughly speaking, in the course of their derivation, the authors bound $\av{|g|^{2}}_{T} $  by separately bounding the absolute value of every term in its Fourier expansion, disregarding the  interference between different terms due to dephasing. 
Each of these terms, which rotate according to $ \exp[i (G_{\alpha}-G_{\beta})T]$ does gets individually dephased, due to the time averaging, but only on a time scale $t_{\alpha \beta}\sim1/(G_{\alpha}-G_{\beta})$. 
The bound in Ref. \cite{Short2012} hence corresponds to the amount of time needed for 
the slowest term in the Fourier sum to average out over time. For many-body systems, the gaps $G_{\alpha}$, and also the differences between different gaps, can be exponentially small in $n$, hence the exponentially long upper bound.

Although this bound is therefore much too large to be a reasonable estimate the equilibration time of most observables, it must be stressed that  one can always construct a specific observable $A$ which saturates it. This is not entirely unexpected, as the bound itself is observable-independent. In fact, in Ref. \cite{Tasaki2013}, Goldstein et al. construct such an observable, by considering a direct sum of banded matrices in the energy basis, each of which has a bandwidth that is exponentially small in the system size. From the dephasing mindset, it is straightforward to understand what is going on in this example.
Since each component of $A$ is banded with an (exponentially) narrow bandwidth, with no coherences between different bands, the dispersion of relevant gaps, $\sigma_{G}$, is exponentially small. In this case, our estimate $\pi/\sigma_{G}$ for the equilibration time becomes exponentially large, since the small difference in angular speeds means that it must take a long time until all points are dephased and isotropically distributed in the complex plane.
Analogously, in  Ref. \cite{Tasaki2013} it is also shown that one can always construct an observable which equilibrates extremely fast, by defining an $A$ which is far from banded, having coherences between vastly different energies. Again, the dephasing picture intuitively explains the reason for such quick equilibration. 

We can also understand some of the results obtained in another approach to the problem of relaxation of many-body systems, namely the study of the survival probability given by the quantum fidelity 
$\mathcal{F}(t)\coloneqq |\bra{\psi(0)}\psi (t) \rangle|^2$, often in situations where the initial state  $\ket{\psi(0)}$ is generated after a sudden displacement (`quantum quench') that brings the system out of equilibrium (see Ref. \cite{ReviewSantos2017} and references therein for a review). 
Note that the quantum fidelity is, up to an additive constant, equivalent to the time evolution of the observable $A=\proj{\psi(0)}$. In this case, then, 
both the energy fluctuations of the initial state
and the bandwidth of the observable written in the energy basis are $\sigma_E$. 
Hence, our considerations within the dephasing picture predict, for local interacting lattice systems, an equilibration time determined by $\sigma_E$, in agreement
with the results of Ref.~\cite{ReviewSantos2017}.

Finally, our outlook and conclusions are also compatible, and in some senses complementary, to recent remarkable results by Reimann et al. \cite{Reimann2016,BalzReimann2017}. In these works, the time signal $g(t)$ in Eqs.(\ref{eq:timesignal}, \ref{eq:thesum}) above is rewritten as $g(t) = c F(t) + \xi(t)$, where (using our notations)
\begin{align}\label{eq:ReimannF}
F(t) = \frac{1}{d_{T}(d_{T}-1)} \sum_{\alpha} e^{\iu G_{\alpha} t},
\end{align}
and c is a constant ($=g(0)$).  Note that $F(t)$ depends only on the gap spectrum, and on the dimension of the system, but not on the initial condition, observable or eigenbasis of the Hamiltonian. It is then proven that, if one averages the time signal over certain ensembles of Hamiltonians with fixed spectra (i.e. varying only their eigenvectors) then both the average value and the standard deviation of $\xi(t)$ become extremely small in the thermodynamic limit. In other words, for \emph{any} fixed initial condition and observable of a quantum system, and also any given fixed energy spectrum, the `typical' time signal will always be of the form $g(t) = c F(t)$, up to negligible error. Comparing with Eq. \eqref{eq:thesum} above, we can see that the effect of the ensemble average is to 
uniformize the different amplitudes $v_{\alpha}$, both in amplitude and in phase. As a result, the `typical' equilibration dynamics described by Eq. (\ref{eq:ReimannF}) becomes a pure dephasing process, i.e, a sum of uniform-length vectors in the complex plane, all initially pointing along the positive real axis, which then rotate at different speeds.

As we have argued above, dephasing/equilibration should then occur on a timescale of order $\pi/\sigma_{G}$. Indeed, this can be seen in the examples worked out in these references. For instance, one situation considered in Ref. \cite{Reimann2016} is a system described by a continuous `microcanonical' density of energy eigenstates, of the form $\rho(E - y) = c e^{- y/ k_{B}T} $, where $y >0$ and $c$ is a normalization constant. It is also assumed that the system's initial state has support restricted to a narrow energy band $[E-\Delta E, E]$. 
In this case $F(t)$ can be calculated exactly; in particular it is shown that, for $\Delta E \gg k_{B}T$, it has the form of a Lorentzian function, 
\begin{align}\label{eq:F}
F(t) =  \frac{1}{1 + (\gamma t )^{2}},
\end{align}
where $\gamma = k_{B} T / \hbar$. 

Let us now analyze this result from the point of view of our approach. First of all, as we have noted in section \ref{sec:equilibration_as_dephasing}, restricting the initial state to a finite energy band, and therefore a finite gap spectrum, ensures that, for all observables, dephasing/equilibration must occur in a finite timescale. 

We can estimate this timescale from the (continuous) gap density for this model:
\begin{align}
\rho_{G}(\omega) = \int \rho(E')\rho(E' +\omega) dE'  =  \frac{\gamma c^{2}e^{- \Delta E/\gamma}}{2} \sinh\left[\frac{\Delta E- |\omega|}{\gamma}\right]  \sim \frac{\gamma c^{2}}{2}e^{- |\omega|/\gamma}. 
\end{align}
where the last equality is valid for $\Delta E \gg k_{B}T$.
Given that here the amplitudes $v_{\alpha}$ of each gap in the time signal are uniform, this expression is also proportional to the gap spectrum (i.e., the Fourier transform of $F(t)$), and we can calculate directly the gap dispersion
\[
\Delta \omega = \frac{\int  \omega^{2} \rho_{G}^{2}(\omega) d \omega}{\int  \rho_{G}^{2}(\omega) d \omega}  = \frac{\gamma}{\sqrt{2}}.
\]
Hence, we estimate an equilibration time $T_{eq} \sim \frac{\pi \sqrt{2}}{\gamma} =  \frac{\pi \sqrt{2} \hbar}{k_{B} T }$. It can be seen from Eq. (\ref{eq:F}),  that, after $t = T_{eq}$, the signal has decreased by about $\sim95\%$ from its initial value, showing that this is indeed a fair estimate of the equilibration timescale. Note that for a time signal governed by Eq. (\ref{eq:F}), the standard deviation $\Delta t$ defined in Eq.~\eqref{eq:time-fluctuations} equals $ 1 / \gamma$, which is also the `half-width at half maximum' of the Lorentzian \footnote{Note that here $\Delta t$ refers to the standard deviation of the \emph{square} of Lorentzian, not that of the Lorentzian itself, which is divergent.}. Hence in this case the uncertainty relation, Eq. \eqref{eq:standardUP} is not far from being saturated, which is the condition we are relying on for our estimate. 

As a final remark, it is also interesting to note that in this case the roles of the time signal and spectrum are exactly reversed from the well-known example of spontaneous decay of an excited atomic or nuclear state: there it is the spectrum that is a Lorentzian, and the time signal that is an exponential decay.
 
\subsection{Equilibration time-scales and level statistics}
\label{subsec:levelstat}
A consequence of the coarse-graining machinery introduced above is that the fine-grained details of the spectrum do not affect the dynamics of the system up to very long time scales.
More formally, the following result states that if two coarse-grained frequency signals are close in the trace norm, then their corresponding time signals must also be point-wise close for any $t<O(\epsilon^{-1})$:

\begin{result}{[Spectrum and dynamics]}\label{result:spectrum-dynamics} Let $g^{(1)}(t)$ and $g^{(2)}(t)$ be two time signals whose $\epsilon$-coarse-grained Fourier transforms satisfy 
\begin{equation} \label{eq:spectrum-dynamics}
\norm{\tilde{g}_{\epsilon}^{(1)}-\tilde{g}_{\epsilon}^{(2)}}_{1}\le\delta_{1}
\end{equation}
for some $\delta_{1} >0$, and let some $0< \delta_{2}<2$ set a distinguishably threshold. Then $g^{(1)}(t)$ and $g^{(2)}(t)$ are indistinguishable up to times $t\le \sqrt{\delta_{2}}/\epsilon$, in the sense that
\begin{equation}
|g^{(1)}(t)-g^{(2)}(t)|< \delta_{1}+\delta_{2}, \label{eq:bound-distinguishability-time-signals}
\end{equation}

\end{result}
\begin{proof} By using twice the triangular inequality, we have 
\begin{equation}
|g^{(1)}(t)-g^{(2)}(t)|\le|g^{(1)}(t)-g_{\epsilon}^{(1)}(t)|+|g_{\epsilon}^{(1)}(t)-g_{\epsilon}^{(2)}(t)|+|g_{\epsilon}^{(2)}(t)-g^{(2)}(t)|\,.
\end{equation}
The first and third terms are analogous and are bounded by Eq.~(\ref{eq:coarse-grained-time-signal}) and the fact that $|g(t)| \leq 1$.
Concerning the second term, we note that, since $\tilde{g}_{\epsilon}(\omega) \in L^{2}$, then 
the uniform continuity statement of functional analysis allows us to write
\begin{equation}
\norm{g_{\epsilon}^{(1)}-g_{\epsilon}^{(2)}}_{\infty}=\sup_{t}|g_{\epsilon}^{(1)}(t)-g_{\epsilon}^{(2)}(t)|\le\norm{\tilde{g}_{\epsilon}^{(1)}-\tilde{g}_{\epsilon}^{(2)}}_{1}\le \delta_{1},
\end{equation}
and putting everything together, we get 
\begin{equation}
|g^{(1)}(t)-g^{(2)}(t)|\le2\left|1-\e^{-\frac{1}{2}\varepsilon^{2}t^{2}}\right|+\delta_{1}.
\end{equation}
Finally, the bound \eqref{eq:bound-distinguishability-time-signals} is respected as long as $1/2 \epsilon^2 t^2\le -\log(1-\delta_2/2)$. The inequality
$x\le -\ln(1-x)$ for $0<x<2$ implies that $1/2 \epsilon^2 t^2\le \delta_2/2 \le -\log(1-\delta_2/2)$. 
Thus, Eq.~\eqref{eq:bound-distinguishability-time-signals} is guaranteed to hold for times
$t\le \sqrt{\delta_2}/\epsilon$.
\end{proof}

Result \ref{result:spectrum-dynamics} shows that two frequency signals that
become very similar once they are $\epsilon$-coarse-grained have indistinguishable dynamics
up to times $\epsilon^{-1}$.
This is particularly relevant in many-body systems where the separation between consecutive energy levels shrinks exponentially in the 
 system size\footnote{This is a consequence of the fact that
in such systems the energy scales extensively (or even polynomially for some long range interactions) with the number of particles, while the dimension of the Hilbert space does so exponentially.}. One can then consider the possibility of two many-body Hamiltonians with qualitatively different \emph{level statistics}, for example one where the distribution of gaps $E_{i+1}-E_i$ between consecutive energy levels follows a Poisson distribution, and another with Wigner-Dyson statistics,
giving rise to time signals that are nevertheless indistinguishable in practice for time up to and beyond the equilibration time. 

In \ref{app:level-statistics} we present an example of this kind. We have estimated the one-norm distance between
two coarse-grained frequency signals evolving according to two Hamiltonians
with identical eigenbases but different level statistics (one Poissonian and the other Wigner-Dyson). For this case we find that
\begin{equation}\label{eq:distance-freq-signals-estimate}
\norm{g_\epsilon^{(1)}-g_\epsilon^{(2)}}_1\leqslant \frac{C}{\epsilon} 
\sqrt{\frac{n^3}{d_{\rm eff}}}
\end{equation}
where $C$ is a numerical constant, $n$ is the system size, 
$\epsilon$ the coarse-graining parameter and $d_{\rm eff}$ the effective dimension.
 
Note now that, assuming that $d_{\rm eff}$ increases exponentially in $n$ ($d_{\rm eff} \sim \exp(cn)$, for some constant $c$),  then choosing $\epsilon\sim \exp(-c n/4)$  in Eq.~\eqref{eq:distance-freq-signals-estimate} and using Result \ref{result:spectrum-dynamics} implies that the dynamics would not be affected up to times $t \sim \exp(c n/4)$.  (Here we are also assuming, as in the discussion following Eq.(\ref{eq:epsilon_condition}),  $\epsilon \gg E_{i+1}-E_i$, which also decrease exponentially. This requires choosing $d_{\rm eff} $ to increase sufficiently slowly ($c$ sufficiently small)).
If these conditions are met, we obtain time signals originating from two Hamiltonians with different level-statistics but that for all practical purposes display the same dynamical behaviour.

Recall now that it is a well-known conjecture that a Poissonian nearest-neighbour gap distribution is a
manifestation of integrability, and Wigner-Dyson statistics are a signature of quantum chaos. Our example seems therefore to show that it is possible for both kinds of Hamiltonian to lead to identical time signals, with identical equilibration times, at least some specific cases.

It is less clear what will happen  in a more realistic example in which the two Hamiltonians $H_1$ and $H_2$ with different level statistics  also have different eigenbases (as is in practice always the case). 
In these situations the two Hamiltonians 
are related by a perturbation $V$ which does not commute with 
the integrable Hamiltonian $H_1$
and where both $H_1$ and $V$ have a locality structure of the type in \eqref{eq:local-hamiltonian}.
In such a scenario, the coarse-grained energy density of the initial state is not affected by the perturbation since it keeps the Hamiltonian local. Thus, any change in the one-norm distance between frequency signals, and thereby in the dynamical behaviour
of the system, must come from a drastic change in the matrix-elements of the observable.
A question that arises beyond the scope of this paper is then how integrability, non-integrability,
and chaos can be identified in the behaviour of the matrix-elements of the observable in the energy basis.

\subsection{The dephasing mindset for quadratic Hamiltonians}
A relevant point to discuss is to which extent the results presented in this paper
are valid for integrable models. 
In this respect, let us focus on quadratic (bosonic or fermionic) Hamiltonians that can be brought to  
a diagonal form $H=\sum_k \varepsilon(k) a^{\dagger}_k a_k$, where $a_k^{(\dagger)}$ are 
the annihilation (creation) operators that fulfil fermionic or bosonic commutation relations 
and $\varepsilon(k)$ is the dispersion relation.
For such systems and quadratic observables, 
the time signal can also be written in the form of Eq.~\eqref{eq:thesum}, where 
the sum now does not run over the gaps of the Hamiltonian, but
over the gaps of the dispersion relation $\varepsilon(k)$.
In Ref.~\cite{Marti2017} this is done in detail for the Caldeira-Leggett model 
(a quadratic system of harmonic oscillators).
In sum, we see that the above formalism can also be applied to quadratic Hamiltonians,
where, roughly speaking, the Hilbert space has been substituted by the space of modes.

\section{Conclusions}

In this work we have argued that equilibration in closed quantum systems 
should be understood as
a process of dephasing of complex numbers in the complex plane.
From this mechanism, we have heuristically estimated the equilibration time-scale
as roughly the inverse of the dispersion of the relevant gaps.
We have seen that, under physically relevant circumstances, 
the equilibration time-scale estimated in this way 
depends at most weakly on the system size, in agreement with
realistic situations. 
Although our argument does not result in a rigorous bound,
we claim that it captures the correct way in which the time-scale depends on the physical properties of the system.  In particular, we have seen that the coherences of the observables 
of interest in the energy basis, $\langle E_{i}|A|E_{j}\rangle$, play a 
fundamental role: in order to attain a finite equilibration time for generic initial states, these 
coherences must become small as $E_{i}-E_{j}$ increases.

We have also observed that the size of the system only plays a role
in the typical size of the fluctuations, but not in the time of equilibration,
and thus small systems fail to equilibrate not because their
equilibration time is large, but because their fluctuations are big.
We illustrate these results with numerical simulations of spins chains.

Finally, we have applied the dephasing mindset to give an intuitive interpretation 
to earlier works on equilibration times. Our results satisfactorily reproduce many particular cases of determining equilibration time scales found in the literature. Of course, further work is still required to put our claims on a more rigorous foundation. For example, we conjecture that in a sufficiently wide range of locally interacting $n$-body systems, the coarse-grained frequency signal $\tilde{g}_{\epsilon}(\omega)$ may itself approach a Gaussian in the limit of large $n$. In this case, the Heisenberg-like uncertainty principle we have been using to heuristically estimate the equilibration timescale would become close to saturated, and would therefore indeed be a bona-fide measure for it.

\subsection*{Acknowledgements}
We would like to thank J.\ Eisert and his co-authors in Ref.~\cite{Wilming2017}
for the coordination in the publication of this manuscript and their work,
and A.\ Molnar and H.\ Wilming for letting us know about the existence of Ref.\ \cite{Arad2016}
which contains a very similar theorem to our Result \ref{result:local-operators}.
TRO and DJ are supported by the Brazilian National Institute for Science and Technology of Quantum Information (INCT-IQ).
TRO also thanks the National Counsel of Technological and Scientific Development (CNPq). CC, ML and AR thank financial support from Spanish MINECO (QIBEQI FIS2016-80773-P, FISICATEAMO FIS2016-79508-P and Severo Ochoa Grant No. SEV-2015-0522), Fundaci\'o Privada Cellex, Generalitat de Catalunya (Grant No. SGR 874, 875, and CERCA Programme) and the European Commission [EQuaM (FP7-ICT-2013-C No.323714), OSYRIS (ERC-2013-AdG No. 339106), QUIC (H2020-FETPROACT-2014 No. 641122), SIQS (FP7-ICT-2011-9 No. 600645)]. 
AR is also supported by the Beatriu de Pin\'os fellowship (BP-DGR 2013) and the CELLEX-ICFO-MPQ research fellowship.

\section*{References}
\bibliography{Equilibration}

\begin{thebibliography}{10}

\bibitem{Gogolin16}
C.~Gogolin and J.~Eisert.
\newblock Equilibration, thermalisation, and the emergence of statistical
  mechanics in closed quantum systems.
\newblock {\em Rep. Prog. Phys}, 79:056001, 2016.

\bibitem{Reimann2007}
P.~Reimann.
\newblock Typicality for generalized microcanonical ensembles.
\newblock {\em Phys. Rev. Lett.}, 99:160404, Oct 2007.

\bibitem{Goldstein2006}
S.~Goldstein, J.~L. Lebowitz, R.~Tumulka, and N.~Zangh{\`\i}.
\newblock Canonical typicality.
\newblock {\em Phys. Rev. Lett}, 96:050403, 2006.

\bibitem{Popescu2006}
S.~Popescu, A.~J. Short, and A.~Winter.
\newblock Entanglement and the foundations of statistical mechanics.
\newblock {\em Nature Phys}, 2:754--758, 2006.

\bibitem{Reimann2008}
P.~Reimann.
\newblock Foundations of statistical mechanics under experimentally realistic
  conditions.
\newblock {\em Phys. Rev. Lett}, 101(19040):3, 2008.

\bibitem{Linden2009}
N.~Linden, S.~Popescu, A.~J. Short, and A.~Winter.
\newblock Quantum mechanical evolution towards thermal equilibrium.
\newblock {\em Phys. Rev. E}, 79:061103, 2009.

\bibitem{Monnai2013}
T.~Monnai.
\newblock Generic evaluation of relaxation time for quantum many-body systems:
  Analysis of the system size dependence.
\newblock {\em Journal of the Physical Society of Japan}, 82(4):044006, 2013.

\bibitem{Brandao2012}
F.~G. S.~L. Brand\~ao, P.~\ifmmode \acute{C}\else
  \'{C}\fi{}wikli\ifmmode~\acute{n}\else \'{n}\fi{}ski, M.~Horodecki,
  P.~Horodecki, J.~K. Korbicz, and M.~Mozrzymas.
\newblock Convergence to equilibrium under a random hamiltonian.
\newblock {\em Phys. Rev. E}, 86:031101, Sep 2012.

\bibitem{Cramer2012}
M~Cramer.
\newblock Thermalization under randomized local hamiltonians.
\newblock {\em New Journal of Physics}, 14(5):053051, 2012.

\bibitem{Masanes2013}
L.~Masanes, A.~J. Roncaglia, and A.~Ac\'{\i}n.
\newblock Complexity of energy eigenstates as a mechanism for equilibration.
\newblock {\em Phys. Rev. E}, 87:032137, Mar 2013.

\bibitem{Malabarba2014}
A.S.L. Malabarba, L.P. Garc{\'\i}a-Pintos, N.~Linden, T.~C. Farrelly, and A.~J
  Short.
\newblock Quantum systems equilibrate rapidly for most observables.
\newblock {\em Physical Review E}, 90(1):012121, 2014.

\bibitem{Goldstein2014}
S.~Goldstein, T.~Hara, and H.~Tasaki.
\newblock The approach to equilibrium in a macroscopic quantum system for a
  typical nonequilibrium subspace.
\newblock {\em arXiv preprint 1402.3380}, 2014.

\bibitem{Goldstein2015}
S.~Goldstein, T.~Hara, and H.~Tasaki.
\newblock Extremely quick thermalization in a macroscopic quantum system for a
  typical nonequilibrium subspace.
\newblock {\em New J. Phys.}, 17:045002, 2015.

\bibitem{Garcia15}
L.~P. Garc{\'\i}a-Pintos, N.~Linden, A.~S.~L. Malabarba, A.~J. Short, and
  A.~Winter.
\newblock Equilibration time scales of physically relevant observables.
\newblock {\em Phys. Rev. X}, 7:031027, 2017.

\bibitem{Short2012}
A.~J. Short and T.~C. Farrelly.
\newblock Quantum equilibration in finite time.
\newblock {\em New J. Phys.}, 14:013063, 2012.

\bibitem{Farrelly2016_QGases}
T.~Farrelly.
\newblock Equilibration of quantum gases.
\newblock {\em New J. Phys.}, 18(7):073014, 2016.

\bibitem{Reimann2016}
P.~Reimann.
\newblock Typical fast thermalization processes in closed many-body systems.
\newblock {\em Nature Comm.}, 7:10821, 2016.

\bibitem{BalzReimann2017}
B.~N. Balz and P.~Reimann.
\newblock Typical relaxation of isolated many-body systems which do not
  thermalize.
\newblock {\em Phys. Rev. Lett.}, 118:190601, May 2017.

\bibitem{ReviewSantos2017}
L.~F. Santos and E.~J. Torres-Herrera.
\newblock Nonequilibrium quantum dynamics of many-body systems.
\newblock {\em arXiv preprint 1706.02031}, 2017.

\bibitem{Tasaki2013}
S.~Goldstein, T.~Hara, and H.~Tasaki.
\newblock On the time scales in the approach to equilibrium of macroscopic
  quantum systems.
\newblock {\em Phys. Rev. Lett.}, 111:140401, 2013.

\bibitem{Srednicki1994}
M.~Srednicki.
\newblock Chaos and quantum thermalization.
\newblock {\em Phys. Rev. E}, 50:901, 1994.

\bibitem{Wilming2017}
H.~Wilming, M.~Goihl, C.~Krumnow, and J.~Eisert.
\newblock Towards local equilibration in closed interacting quantum many-body
  systems.
\newblock {\em arxiv preprint 1704.06291}, 2017.

\bibitem{Arad2016}
I.~Arad, T.~Kuwahara, and Z.~Landau.
\newblock Connecting global and local energy distributions in quantum spin
  models on a lattice.
\newblock {\em J. Stat. Mech.}, 3:033301, 2016.

\bibitem{Short2011}
A.~J. Short.
\newblock Equilibration of quantum systems and subsystems.
\newblock {\em New J. Phys.}, 13:053009, 2011.

\bibitem{Reimann2010}
P.~Reimann.
\newblock Canonical thermalization.
\newblock {\em New J. Phys.}, 12(5):055027, 2010.

\bibitem{Fonda}
L.~Fonda, G.~C. Ghirardi, and A.~Rimini.
\newblock Decay theory of unstable quantum systems.
\newblock {\em Reports on Progress in Physics}, 41(4):587, 1978.

\bibitem{Robinett2004}
R.~W. Robinett.
\newblock Quantum wave packet revivals.
\newblock {\em Physics Reports}, 392:1--119, 2004.

\bibitem{Eberly1980}
J.~H. Eberly, N.~B. Narozhny, and J.~J. Sanchez-Mondragon.
\newblock Periodic spontaneous collapse and revival in a simple quantum model.
\newblock {\em Phys. Rev. Lett.}, 44:1323, 1980.

\bibitem{Narozhnyetal}
J.~H. Eberly N.~B. Narozhny and J.~J. Sanchez-Mondragon.
\newblock Coherence versus incoherence: Collapse and revival in a simple
  quantum model.
\newblock {\em Phys. Rev. A}, 23:236, 1981.

\bibitem{AllenEberlyBook}
L.~Allen and J.~H. Eberly.
\newblock {\em Optical Resonance and Two-Level Atoms}.
\newblock Dover Publications, 1987.

\bibitem{PitaevskiiBook}
E.~M. Lifshitz and L.~P. Pitaevskii.
\newblock {\em Statistical Physics: Theory of the Condensed State (Course of
  Theoretical Physics vol. 9)}.
\newblock Pergamon Press, 1958.

\bibitem{BaymPethickBook}
G.~Baym and Christopher Pethick.
\newblock {\em Landau Fermi-Liquid Theory: Concepts and Applications}.
\newblock Wiley-VCH, 2004.

\bibitem{Polkovnikov11}
A.~Polkovnikov, K.~Sengupta, A.~Silva, and M.~Vengalattore.
\newblock Colloquium: Nonequilibrium dynamics of closed interacting quantum
  systems.
\newblock {\em Rev. Mod. Phys}, 83:863, 2011.

\bibitem{Stein2003}
E.~M. Stein and R.~Shakarchi.
\newblock {\em Fourier Analysis: an Introduction}.
\newblock Princeton University Press, 2015.

\bibitem{Eberlein1949}
W.~F. Eberlein.
\newblock Abstract ergodic theorems and weak almost periodic functions.
\newblock {\em Trans. Amer. Math. Soc}, 67(1):217--240, 1949.

\bibitem{Corduneanu2009}
C.~Corduneanu.
\newblock {\em Almost Periodic Oscillation and Waves}.
\newblock Springer, 2009.

\bibitem{Brandao2015b}
F.~G. S.~L. Brand{\=a}o and M.~Cramer.
\newblock Equivalence of statistical mechanical ensembles for non-critical
  quantum systems.
\newblock {\em arXiv preprint 1502.03263}, 2015.

\bibitem{Dalessio2016}
L.~D'Alessio, Y.~Kafri, A.~Polkovnikov, and M.~Rigol.
\newblock From quantum chaos and eigenstate thermalization to statistical
  mechanics and thermodynamics.
\newblock {\em Adv. Phys.}, 65:239--362, 2016.

\bibitem{Beugeling15}
M.~Haque, W.~Beugeling, and R.~Moessner.
\newblock Off-diagonal matrix elements of local operators in many-body quantum
  systems.
\newblock {\em Phys. Rev. E}, 91:012144, 2015.

\bibitem{Miranda2011}
Y.~M. Miranda and G.~Slade.
\newblock The growth constants of lattice trees and lattice animals in high
  dimensions.
\newblock {\em Electron. Commun. Probab.}, 16:129, 2011.

\bibitem{Marti2017}
M.~Perarnau-Llobet, H.~Wilming, A.~Riera, R.~Gallego, and J.~Eisert.
\newblock Fundamental corrections to work and power in the strong coupling
  regime.
\newblock {\em arXiv preprint 1704.05864}, 2017.

\bibitem{Hastings2006}
M.~B. Hastings and T.~Koma.
\newblock Spectral gap and exponential decay of correlations.
\newblock {\em Communications in Mathematical Physics}, 265(3):781--804, Aug
  2006.

\bibitem{Kliesch2014}
M.~Kliesch, C.~Gogolin, M.~J. Kastoryano, A.~Riera, and J.~Eisert.
\newblock Locality of temperature.
\newblock {\em Phys. Rev. X}, 4:031019, 2014.

\bibitem{Penrose1994}
M.~Penrose.
\newblock Self-avoiding walks and trees in spread-out lattices.
\newblock {\em J. Stat. Phys.}, 77:3, 1994.

\end{thebibliography}
\bibliographystyle{unsrt}

\newpage
\appendix

\section{Energy density and effective dimension of the initial state for local Hamiltonians}
\label{sec:local-Ham}

\subsection{Energy uncertainty of the initial state}

A standard way to bring quantum systems out of equilibrium is to initialize
the system as the ground state of a local Hamiltonian, and change the Hamiltonian 
of the system sufficiently
fast such that the state is kept
unchanged. 
This procedure is called a quantum \emph{quench} \cite{Polkovnikov11}.

It is well known that the ground state of local Hamiltonians exhibit a clustering
of correlations \cite{Hastings2006}. In fact, correlations between observables supported at different lattice points decay algebraically with
the distance for critical systems (gapless Hamiltonians) and exponentially
for systems off criticality (gapped Hamiltonians). 

Consider a local Hamiltonian of the type (\ref{eq:local-hamiltonian})
and a quantum state $\ket{\psi}\in\mathcal{H}$ with either exponentially
or algebraically decaying correlations such that 
\begin{equation}
|\av{h_{u}h_{v}}-\av{h_{u}}\av{h_{v}}|\le\frac{c}{d(u,v)^{D+1}},\label{eq:decaying-correlations}
\end{equation}
where $c>0$, $d(u,v)$ is the graph distance between the edges $u$
and $v$, and $D$ is the spatial dimension in which the graph can
be embedded. Then, the variance of the energy distribution is upper
bounded by $\var_{\psi}(H_{n})\le c'n$ with $c'>0$.

By using (\ref{eq:decaying-correlations}), it is easy
to see that 
\begin{equation}
\sum_{v\in E}\left(\av{h_{u}h_{v}}-\av{h_{u}}\av{h_{v}}\right)\le c'\hspace{0.6cm}\forall\ \ u\,.\label{eq:decaying-correlations2}
\end{equation}
Hence, writing the variance in terms of the local terms of the Hamiltonian,
and using the previous bound, one gets 
\begin{equation}
\textrm{var}_{\psi}(H_{n})=\av{H^{2}}_{\psi}-\av{H}_{\psi}^{2}=\sum_{u,v\in E}\left(\av{h_{u}h_{v}}_{\psi}-\av{h_{u}}_{\psi}\av{h_{v}}_{\psi}\right)\le cn\, .
\end{equation}

\subsection{Effective dimension}

The effective dimension tells us how many eigenstates of the Hamiltonian
contribute in the superposition of the initial state. In the previous
section we have argued that the energy uncertainty of the initial
state scales with $\sqrt{n}$ in the case where the system is brought out of equilibrium with a global quench, and is independent of the system size in the case where the system suffers a local quench.

Local Hamiltonians have an energy range that scales linearly in the
system size while the dimension of the Hilbert space does so exponentially.
This implies that the density of states scales exponentially, and
so does the effective dimension.

\section{Proof of Result \ref{result:coarse-grained-frequency-signal}: Coarse-grained frequency signal.}
\label{app:proof-freq-signal}
In this section we provide the proof of Result \ref{result:coarse-grained-frequency-signal}.
The coarse grained Fourier transform of the time signal
naturally decomposes 
\begin{equation}
\tilde{g}_{\epsilon}(\omega)=\sum_{\alpha}v(G_{\alpha})h_{\epsilon}(\omega-G_{\alpha})+\sum_{\alpha}\delta v_{\alpha}h_{\epsilon}(\omega-G_{\alpha})\,.\label{eq:g_omega-decomposition}
\end{equation}
The second term in the right hand side is a random variable with variance
given by 
\begin{equation}
\delta v_{\epsilon}(\omega)^{2}:=\textrm{var}\left(\sum_{\alpha}\delta v_{\alpha}h_{\epsilon}(\omega-G_{\alpha})\right)=\sum_{\alpha}h_{\epsilon}^{2}(\omega-G_{\alpha})\textrm{var}\left(\delta v_{\alpha}\right)=\sum_{\alpha}h_{\epsilon}^{2}(\omega-G_{\alpha})\gamma^{2}(G_{\alpha})
\nonumber
\end{equation}
By means of the density of gaps, the variance of the fluctuation term
in Eq.~\eqref{eq:g_omega-decomposition} becomes 
\begin{equation}
\delta v_{\epsilon}(\omega)^{2}=\frac{\sqrt{\pi}}{\epsilon}\int\dd\omega'N_{\epsilon/\sqrt{2}}(\omega-\omega')\rho(\omega')\gamma(\omega')\,,
\end{equation}
where we have taken into account that $h_{\epsilon}(\omega)^{2}=\sqrt{\pi}\epsilon^{-1}N_{\epsilon/\sqrt{2}}(\omega)$.
Now we use the fact that $\gamma(\omega)$ is almost constant in an
interval $\epsilon$, i.~e.\ the Lipshitz constant $K$ of $\gamma$
is such that $K\epsilon\ll1$. The idea is to approximate 
$\delta v_{\epsilon}(\omega)^{2}$ by
\begin{equation}
\delta v_{\epsilon}(\omega)^{2}\simeq \frac{\sqrt{\pi}}{\epsilon}\gamma^{2}(\omega)\int\dd\omega'N_{\epsilon/\sqrt{2}}(\omega-\omega')\rho(\omega')=\frac{\sqrt{\pi}}{\epsilon}\gamma^{2}(\omega)\rho_{\epsilon/\sqrt{2}}(\omega)\,,\label{eq:v-fluctuations}
\end{equation}
More precisely, such approximation has 2 steps. The first step is to restrict the domain of integration
around $\omega$, where the Gaussian $N_{\epsilon/\sqrt{2}}(\omega-\omega')$ is centred. 
The error of such step can be upper bounded by
\begin{equation}
\left|\delta v_{\epsilon}(\omega)^{2}-\frac{\sqrt{\pi}}{\epsilon}\int_{\omega-\Delta}^{\omega+\Delta}\dd\omega'N_{\epsilon/\sqrt{2}}(\omega-\omega')\rho(\omega')\gamma(\omega')\right|
\le \frac{\sqrt{\pi}}{\epsilon}\rho_{\epsilon/\sqrt{2}}(\omega)\alpha \e^{-2\Delta^2/\epsilon^2}
\end{equation}
where we have used the Chernoff bound of the Gaussian distribution $\int_{x}^\infty \dd y N_\sigma(y)\le 2 \e^{-x^2/\sigma^2}$ and $\alpha>0$ is a constant that depends on the density of gaps.

The second step is considering the worst case in the variation of  $\gamma(\omega)$ within the interval
of integration $[\omega-\Delta,\omega+\Delta]$.
To do so, let us consider two real positive functions $a(x)$ and $b(x)$, 
where $b(x)$ has Lipschitz constant $K$.
Then, by using trivial calculus and the definition of Lipschitz constant, we get
\begin{equation}
\int_{\omega- \Delta}^{\omega+\Delta} \dd x a(x) b(x) 
\le \left(\max_{x \in [\omega- \Delta,\omega+\Delta]} b(x)\right) 
\int_{\omega- \Delta}^{\omega+\Delta} \dd x a(x) 
\le \left( b(\omega)+K\Delta \right) 
\int_{\omega- \Delta}^{\omega+\Delta} \dd x a(x)\, . \nonumber
\end{equation}
A straight forward application of the former statement leads to
\begin{equation}
\int_{\omega- \Delta}^{\omega+\Delta}\dd\omega'
\gamma^{2}(\omega')N_{\epsilon/\sqrt{2}}(\omega-\omega')\rho(\omega')
-\gamma^{2}(\omega)\int_{\omega- \Delta}^{\omega+\Delta}\dd\omega'N_{\epsilon/\sqrt{2}}(\omega-\omega')\rho(\omega')
\le \rho_{\epsilon/\sqrt{2}}(\omega) K  \Delta\, .\nonumber
\end{equation}
By putting everything together, we get
\begin{equation}
\left|\delta v_{\epsilon}(\omega)^{2}-\frac{\sqrt{\pi}}{\epsilon}\gamma^{2}(\omega)\rho_{\epsilon/\sqrt{2}}(\omega)\right|\le \frac{\sqrt{\pi}}{\epsilon}\rho_{\epsilon/\sqrt{2}}(\omega) \left(2\alpha \e^{-2\Delta^2/\epsilon^2}+ K \Delta\right)
\end{equation}
Finally, an optimization over $\Delta$ gives
\begin{equation}\label{eq:error-bound-fluctuatuions}
\left|\delta v_{\epsilon}(\omega)^{2}-\frac{\sqrt{\pi}}{\epsilon}\gamma^{2}(\omega)\rho_{\epsilon/\sqrt{2}}(\omega)\right|\le \alpha'\epsilon^{-1} \rho_{\epsilon/\sqrt{2}}(\omega) K \epsilon 
\end{equation}
where $\alpha'>0$ contains all the numerical factors.
Equation \eqref{eq:error-bound-fluctuatuions} implies a relative error in the
of the order $K \epsilon$.

The standard deviation $\delta v_{\epsilon}(\omega)$ sets the order
of magnitude for the fluctuations of the random variable. We will see that it is negligible
in comparison with the first term in Eq.~\eqref{eq:g_omega-decomposition}.
Such a term can also be written in terms of the density of gaps $\rho(\omega)$
\begin{equation}
R_{\epsilon}(\omega):=\sum_{\alpha}v(G_{\alpha})h_{\epsilon}(\omega-G_{\alpha})=\int\dd\omega'\rho(\omega')v(\omega')h_{\epsilon}(\omega-\omega')\label{eq:relevant-gaps-R}
\end{equation}
The function $R_{\epsilon}(\omega)$ can be shown to be close to $\sqrt{2\pi}\ v(\omega)\rho_{\epsilon}(\omega)$. More precisely, by following an analogous argument of the one above,
it can be proven that
\begin{equation}\label{eq:approx_R}
\left|R_{\epsilon}(\omega)-\sqrt{2\pi}\ v(\omega)\rho_{\epsilon}(\omega)\right| \le \beta \rho_{\epsilon}(\omega) K \epsilon
\end{equation}
where $\beta>0$ is an order one factor.

We are now ready to bound the difference $\tilde{g}_{\epsilon}(\omega)-\sqrt{2\pi} v(\omega)\rho_{\epsilon}(\omega)$.
By using Eq.~\eqref{eq:g_omega-decomposition}, definition \eqref{eq:relevant-gaps-R} 
and the triangular inequality, we get
\begin{equation}
\left|\tilde{g}_{\epsilon}(\omega)-\sqrt{2\pi} v(\omega)\rho_{\epsilon}(\omega)\right|\le\left|R_{\epsilon}(\omega)-\sqrt{2\pi} v(\omega)\rho_{\epsilon}(\omega)\right|+\left|\sum_{\alpha}\delta v_{\alpha}h_{\epsilon}(\omega-G_{\alpha})\right| 
\end{equation}
Finally, from Eqs.~\eqref{eq:v-fluctuations} and \eqref{eq:approx_R},
the error is bounded by
\begin{equation}
\left|\tilde{g}_{\epsilon}(\omega)-\sqrt{2\pi} v(\omega)\rho_{\epsilon}(\omega)\right|\le
\rho_{\epsilon}(\omega)\left(\beta  K \epsilon+
\frac{\pi^{1/4}\gamma(\omega)}{\sqrt{\epsilon \rho_{\epsilon}(\omega)}}\right)\, .
\end{equation}
Here, we have used that $\rho_{\epsilon/\sqrt{2}}(\omega)\le \rho_{\epsilon/\sqrt{2}}(\omega)(1+ K \epsilon)$ where $K$ also upper-bounds the Lipshitz constant of $\rho_\epsilon(\omega)$.

\section{Proof of Result \ref{result:local-operators}: Local operators are banded in the energy basis}
\label{app:proof-local-operators}
In this section we prove result \ref{result:local-operators}.
We first introduce the imaginary time evolution of the observable
\begin{equation}\label{eq:imaginary-time-evolution}
A_x(\beta):=\e^{\beta H} A_x \e^{-\beta H}\, .
\end{equation}
By Taylor expanding of the exponentials, it can be written as
\begin{equation}
A_x(\beta)=A_x+\sum_{m=1}^\infty \frac{\beta^m}{m!}[H,\ldots[H,A_x]\ldots]\, ,
\end{equation}
where we have used that $[H,\ldots[H,A_x]\ldots]=\sum_{k=0}^m  \binom{m}{k} H^k A_x (-H)^{m-k}$.

We use now the locality of the Hamiltonian given by its underlying graph $(V,\mathcal{E})$.
In order to do so and
for the sake of clarity, let us introduce some notation. 
We refer to edges $\lambda \in \mathcal{E}$ as \emph{letters}, to the edge set $\mathcal{E}$ as an \emph{alphabet}, and call the sequences of edges as \emph{words}. 
For any \emph{sub-alphabet} $F \subset \mathcal{E}$, we denote by 
$F^l$  
the set of words with letters in $F$ and length $l$, where the length $|w|$ of a word $w$ is the total number of letters it contains.
In turn, the set of words with letters in $F$ and arbitrary length $l$ is defined as 
$F^\ast := \bigcup_{l=0}^\infty F^l$.
The size of a sub-alphabet $F$ is denoted by $|F|$ and corresponds to the number of letters
it contains.
We call a sub-alphabet $F$ an animal if it forms a set of \emph{connected} edges,
and  a word \emph{connected} or \emph{cluster} $c \in \mathcal{E}^\ast$   
if the set of letters in $c$ is an animal, i.e., connected.
That is, clusters are connected sequences of edges where the edges can also occur multiple times, while animals are connected sub-graphs (sets of connected edges without order or repetition). 
For $w \in E^\ast$ and any sub-alphabet $G\subset \mathcal{E}$, we write $G \subset w$ if every letter in $G$ also occurs in $w$. 
When a sub-alphabet $F$ contains at least one edge adjacent to a vertex $x$, we denote it by
$F\ni x$.

With the above notation, the order $n$ commutator can be written as
\begin{equation}
[H,\ldots[H,A_x]\ldots]=\sum_{w\in \mathcal{E}^m} [h_{w_1},\ldots [h_{w_m},A_x]\ldots]
\end{equation}
where $w_k$ is the $k$-th letter of the word $w$ and $h_{w_k}$ the corresponding 
Hamiltonian term.
Note that the only way for
the $w$-commutator $[h_{w_1},\ldots [h_{w_n},A_x]\ldots]$ to be nonzero
is if $w$ forms a single cluster with at least one of its letters
adjacent to the vertex $x$.
More formally,
\begin{equation}
[H,\ldots[H,A_x]\ldots]=\sum_{\substack{F\subset \mathcal{E}:\\ |F|\le n \\ F\ni x}}\hspace{2mm}\sum_{\substack{w\in F^n: \\F \subset w}} [h_{w_1},\ldots [h_{w_n},A_x]\ldots]
\end{equation}
where the first sum runs over all the animals $F$ that contain at least an edge adjacent to the vertex $x$ and are smaller or equal than $n$.
Then, the imaginary time evolution of the observable can be written as
\begin{equation}
A_x(\beta)=A_x+\sum_{\substack{F\subset \mathcal{E}:\\ F\ni x}}\hspace{2mm} 
\sum_{\substack{w\in F^*:\\ F \subset w}}\frac{\beta^{|w|}}{|w|!} [h_{w_1},\ldots [h_{w_{|w|}},A_x]\ldots]\, .
\end{equation}
We can upper bound the operator norm of $A_x(\beta)$ as
\begin{equation}
\norm{A_x(\beta)}\le \norm{A_x}\left(1+ 
\sum_{\substack{F\subset \mathcal{E}:\\ F\ni x}}\hspace{2mm} 
\sum_{\substack{w\in F^*:\\ F \subset w}}\frac{(J\beta)^{|w|}}{|w|!}\right)
\end{equation}
where we have used that for any operator $O$, $\norm{[h_{w_k},O]}\le \norm{h_{w_k}}\norm{O}$, 
whose iteration implies $\norm{[h_{w_1},\ldots [h_{w_{|w|}},A_x]\ldots]}\le \norm{A_x}J^{|w|}$.
We recall now Lemma 5 from Ref.~\cite{Kliesch2014} which states that
\begin{equation}
\sum_{w\in F^*: F \subset w}\frac{|\beta J|^{|w|}}{|w|!}=(\e^{|\beta J|}-1)^{|F|}
\end{equation}
and implies 
\begin{equation}
\norm{A_x(\beta)}\le \norm{A_x}\left(1+ \sum_{\substack{F\subset \mathcal{E}:\\ F\ni x}} (\e^{|\beta J|}-1)^{|F|} \right)\, .
\end{equation}
By decomposing the sum over all animals which contain $x$ according to their size, we get
\begin{equation}\label{eq:norm-A(beta)aux}
\norm{A_x(\beta)}\le \norm{A_x} \left(1+\sum_{l=1}^\infty 
\sum_{\substack{F\subset \mathcal{E}:\\  |F|=l \\ F\ni x}} (\e^{\beta J}-1)^{|F|}
\right),
\end{equation}
The number of lattice animals of size $l$ is upper-bounded by means of the so called \emph{lattice animal constant} $\alpha$.
Given a graph $(V,\mathcal{E})$ and denoting by $a_l$ the number of lattice animals of size $l$ containing the fixed vertex $x$,
then the \emph{animal constant} $\alpha$ is the smallest constant satisfying
\begin{equation}\label{eq:lattice-constant}
  a_l:=\sum_{\substack{F\subset \mathcal{E}:\\  |F|=l \\ F\ni x}} 1 \leq \alpha^l .
\end{equation}
Regular lattices have finite animal constants\cite{Penrose1994}. 
For example, the animal constant of a $D$-dimensional cubic lattice can be bounded as $\alpha \leq 2\,D\,\e$ (Lemma~2 in Ref.~\cite{Miranda2011}), where $\e $ is Euler's number. 

Equation \eqref{eq:lattice-constant} allows us to bound \eqref{eq:norm-A(beta)aux} as
\begin{equation}
\norm{A_x(\beta)}\le \norm{A_x} \sum_{l=0}^\infty \left(\alpha(\e^{\beta J}-1)\right)^l,
\end{equation}
which is a geometric series with common ratio $\alpha(\e^{2\beta J}-1)$.
In order for the series to converge we require that $\beta$ is such that
$\alpha(\e^{\beta J}-1)<1$, for which
\begin{equation}\label{eq:bound_A(beta)}
\norm{A_x(\beta)}\le \norm{A_x} \frac{1}{1-\alpha(\e^{\beta J}-1)}\, .
\end{equation}

We compute the absolute value of the matrix-element $\bra{E_{i}}\cdot\ket{E_{j}}$
in Eq.~\eqref{eq:imaginary-time-evolution}
\begin{equation}
\e^{\beta(E_{i}-E_{j})}|\bra{E_{i}}A_{x}\ket{E_{j}}|=|\bra{E_{i}}A_x(\beta)\ket{E_{j}}|\le \norm{A_x(\beta)}\,.
\end{equation}
and, together with Eq.~\eqref{eq:bound_A(beta)}, it implies that
\begin{equation}\label{eq:upper-bound-mat-elementA}
|\bra{E_{i}}A_{x}\ket{E_{j}}|\le \norm{A_x} \frac{\e^{- \beta J (E_{i}-E_{j})/J}}{1-\alpha(\e^{\beta J}-1)}
\end{equation}
for any $\beta$ such that $\alpha(\e^{\beta J}-1)<1$.

In order to get an explicit bound independent of $\beta$, we optimize over $\beta$. 
Given some energy difference $E_{i}-E_{j}$, we look for the
$\beta$ that minimizes the upper bound of Eq.~\eqref{eq:upper-bound-mat-elementA}.
To do so, it is useful to rewrite the bound in terms of a new parameter $z=\e^{\beta J}$
such that $|\bra{E_{i}}A_{x}\ket{E_{j}}|\le \norm{A_x}f(z)$ with
\begin{equation}
f(z):=\frac{1}{\alpha(1+\alpha^{-1}-z)z^{\omega}}
\end{equation}
and $\omega=(E_i-E_j)/J$.
By means of an optimization over $z$, 
we find the minimum at $z=\frac{\omega}{\omega+1}\frac{\alpha+1}{\alpha}$
which fulfils the convergence condition $1<z<1+\alpha^{-1}$ as long as $\omega>\alpha$.
The minimum value of $f(z_*)$ becomes
\begin{equation}
f(z_*)=\frac{\omega+1}{\alpha+1}\left(\frac{\omega}{\omega+1}\right)^{-\omega}\left(1+\alpha^{-1}\right)^{-\omega}
\le \frac{\omega+1}{\alpha+1}\e\left(1+\alpha^{-1}\right)^{-\omega}
\end{equation}
where the inequality comes from the fact that $\left(\omega/(\omega+1)\right)^{-\omega}\le \e$.
The matrix element $|\bra{E_{i}}A_{x}\ket{E_{j}}|$ is then bounded by
\begin{equation}
|\bra{E_{i}}A_{x}\ket{E_{j}}|\le \norm{A_x}\e^{\log\left(\frac{\e(E_i-E_j)}{J(1+\alpha)}\right) -c (E_{i}-E_{j})/J}
\end{equation}
where $c=\log(1+\alpha^{-1})$ is the decay rate.

\section{One-norm distance between coarse-grained frequency signals of two systems which only differ in the level statistics}
\label{app:level-statistics}
Two Hamiltonians $H^{(1)}$ and $H^{(2)}$ with identical eigenstates but eigenvalues with different level statistics give rise to time signals for equal initial states and observables given by
\begin{equation}
g^{(1)}(t)= \sum_\alpha v_\alpha \e^{\iu G_\alpha^{(1)} t}\, ,
\end{equation}
and analogously for $g^{(2)}(t)$. Here, $G_\alpha^{(1)}$ are the gaps associated
the spectrum of $H_1$, and the relevances $v_\alpha$ are the same for both systems.
These time-signals have Fourier transforms
\begin{equation}
\tilde g^{(1)}(\omega)= \sum_\alpha v_\alpha \delta (\omega- G_\alpha^{(1)})
\end{equation}
that once coarse grained read
\begin{equation}\label{eq:cg-frequency-signal-(1)}
\tilde g_\epsilon^{(1)}(\omega)= \sum_\alpha 
v_\alpha h_\epsilon (\omega- G_\alpha^{(1)})\, .
\end{equation}
The one-norm distance between the frequency signals $\tilde g_\epsilon^{(1)}$ and $\tilde g_\epsilon^{(2)}$ is defined by
\begin{equation}
\normone{\tilde g_\epsilon^{(1)}-\tilde g_\epsilon^{(2)}}\coloneqq\int \dd \omega |g^{(1)}(\omega)-g^{(2)}(\omega)|\, .
\end{equation}
Let us now introduce the function
\begin{equation}
u_\delta (\omega)=2 \epsilon h_\epsilon(\omega)h_\epsilon(\delta)\sinh\left( \frac{\delta}{\epsilon} \frac{\omega}{\epsilon}\right)\, .
\end{equation}
By plugging \eqref{eq:cg-frequency-signal-(1)}, we can write the difference in frequency signals as
\begin{equation}
g^{(1)}(\omega)-g^{(2)}(\omega)=\sum_\alpha v_\alpha \ u_{\delta G_\alpha}(\omega-\bar G_\alpha)\, ,
\end{equation}
where we have used that 
$u_{\delta G_\alpha}(\omega-\bar G_\alpha)=h_\epsilon(\omega-G_\alpha^{(1)})-h_\epsilon(\omega-G_\alpha^{(2)})$
with $\delta G_\alpha = G_\alpha^{(1)}-G_\alpha^{(2)}$ and $\bar G_\alpha = (G_\alpha^{(1)}+G_\alpha^{(2)})/2$.

In order to understand how the variables $\bar G_\alpha$ and $\delta G_\alpha$ behave,
let us introduce the spectrum of $H_1$ (with Poisson level statistics)
by $\{E_k^{(1)}\}_{k=1}^{d_E}$, 
and the one of $H_2$ (with Wigner-Dyson level statistics) 
by $\{E_k^{(2)}\}_{k=1}^{d_E}$.
The energy levels can be iteratively constructed as
\begin{equation}
E_{k+1}=E_k+s_k
\end{equation}
where $s_k$ $\forall \ k$ are i.~i.~d.\ random variables
sampled from a Poisson distribution in the integrable case and from a Wigner-Dyson distribution in the chaotic case.
Both distributions are assumed to have the same mean $\mu_s$ and standard deviation 
$\sqrt{\mu_s}$.
The gaps read then
\begin{equation}
G_\alpha  = E_j - E_i = \sum_{k=i}^{j-1} s_k
\end{equation}
where we have assumed $i<j$.
For $i\ll j$ we can apply the central limit theorem and obtain that
the gaps are random variables 
\begin{equation}
G_\alpha \sim N(\mu_s(j-i),\sqrt{\mu_s(j-i)})\, .
\end{equation}
where $N(\mu,\sigma)$ is the Gaussian distribution with mean $\mu$ and standard deviation $\sigma$.
The difference of gaps generated by different probability distributions 
\begin{equation}
\delta G_\alpha\coloneqq G_\alpha^{(1)}-G_\alpha^{(2)}, \sim N(0, \sqrt{2(j-i)\mu_s})\, .
\end{equation}
Note the standard deviation increasing with the separation of the energy levels.
This is indeed the behaviour of $\delta G_\alpha$ if both $s_k^{(1)}$ and $s_k^{(2)}$ are independent random variables. 
However, in our example the spectrum $E_k^{(2)}$ is built from $E_k^{(1)}$ with the
single goal of changing its level-statistics. This can be made shifting the energy levels by an amount independent of the separation between them.
One way to achieve so is to group the energy levels $\{E_k^{(1)}\}_{k=1}^d$ 
in different sets $\{E_k^{(1)}\}_{k=j\,L}^{(j+1)L-1}$ of consecutive $L$ energy levels, where $j$ labels the different sets. Then, the first and the last energy levels of every set are kept fixed 
and the other energy levels are shifted according to the other level-statistics. 
In such a case, the random variable $\delta G_\alpha$ is bounded by
\begin{equation}
\delta G_\alpha \leqslant K\sqrt{\mu_S}
\end{equation}
with $K=\sqrt{2 L}$ a constant independent of the system size.

One strategy to bound the one point distance between the coarse-grained frequency signals would be to use the triangular inequality
as follows
\begin{equation}
|g_\epsilon^{(1)}(\omega)-g_\epsilon^{(2)}(\omega)|\leqslant  \sum_\alpha  |v_\alpha||u_{\delta G_\alpha}(\omega-\bar G_\alpha)|\, .
\end{equation}
and the one norm distance
\begin{equation}
\norm{g_\epsilon^{(1)}-g_\epsilon^{(2)}}_1\le \sum_\alpha  |v_\alpha|\int \dd \omega |u_{\delta G_\alpha}(\omega-\bar G_\alpha)|\, .
\end{equation}

In order to estimate the scaling of the $|v_\alpha|$ in the system size, let us
consider that there are $d_{\rm eff}$ energy levels with a non-negligible occupation in the
initial state.
Thus, 
let us note that
\begin{equation}
\sum_\alpha |v_\alpha|^2 \sim d_{\rm eff}^2 |v_\alpha|^2 \leqslant \frac{1}{d_{\rm eff}}\, ,
\end{equation} 
and therefore $|v_\alpha|\sim d_{\rm eff}^{-3/2}$.

By splitting the integrals and a convenient change of variables, 
\begin{equation}\label{eq:bound-on-integral}
\begin{split}
\int \dd \omega |u_\sigma(\omega)|&= 2   
\left|\int_{-\infty}^{\frac{\sigma}{2}} \dd \omega 
h_\epsilon (\omega)
-\int_{-\infty}^{-\frac{\sigma}{2}}
 \dd \omega h_\epsilon (\omega)\right|\\
& =
2 \left|\int_{-\frac{\sigma}{2}}^{\frac{\sigma}{2}} \dd \omega 
h_\epsilon (\omega)
\right|
\leqslant \frac{2 \sigma}{\epsilon}\, .
\end{split}
\end{equation}
Note that for $\delta G_\alpha \leqslant K \sqrt{\mu_s}$ with $\mu_s \sim n / d_{\rm eff}$,
the individual terms of the sum behave as $K \sqrt{n}/(\epsilon d_{\rm eff}^2)$.
However, the sum over $\alpha$ contains $d_{\rm eff}^2$ many terms, and therefore the bound
becomes useless.
The reason for that is that, as it happened for the time signal, the quantity $|g^{(1)}(\omega)-g^{(2)}(\omega)|$ is small due to the cancellations between its different contributions 
$u_{\delta G_\alpha}(\omega)$.

Instead, let us exploit interference by applying the triangular inequality as follows
\begin{equation}
|g^{(1)}(\omega)-g^{(2)}(\omega)|\leqslant \sum_k \left|\sum_{\bar G_\alpha\in I_k} v_\alpha u_{\delta G_\alpha}(\omega-\bar G_\alpha)\right|
\end{equation}
where $I_k\coloneqq [k \delta \omega, (k+1)\delta \omega)$. 

For given $k$, the contribution to the one-norm behaves as
\begin{equation}\label{eq:interference}
\int \dd \omega\left|\sum_{\bar G_\alpha\in I_k} v_\alpha u_{\delta G_\alpha}(\omega-\bar G_\alpha)\right|
\sim \sqrt{d_k} \av{v_\alpha}_k \int \dd \omega |u_{c\sqrt{\mu_s}}(\omega)| 
\leqslant \frac{K}{\epsilon}\sqrt{\frac{d_k \mu_s }{d_{\rm eff}^{3}}}
\end{equation}
where $d_k\coloneqq |\{\bar G_\alpha \in I_k\}|$ is the number of gaps in the interval $I_k$ and we have used $|v_\alpha|\sim d_{\rm eff}^{-3/2}$ and Eq.~\eqref{eq:bound-on-integral}.
Note that in \eqref{eq:interference} and due to interference between 
different $u_{\delta G_\alpha}(\omega-\bar G_\alpha)$
the scaling with the number of gaps within an interval $I_k$ is 
$\sqrt{d_k}$ and not $d_k$. This behaviour has been inferred from numerical simulations
and can be understood by means of the central limit theorem.

Finally, let us note that the amount of gaps in an interval $I_k$ scales as $d_k\sim C_k d_{\rm eff}^2$ where $d_{\rm eff}^2$ is the total number of gaps, $n$ the the system size, and 
$C_k$ the fraction of gaps in the interval $I_k$. 
Thus, adding all the contributions of every interval
\begin{equation}
\norm{g_\epsilon^{(1)}-g_\epsilon^{(2)}}_1\leqslant \frac{C}{\epsilon} 
\sqrt{\frac{n^3}{d_{\rm eff}}}
\end{equation}
where we have used that the average level spacing $\mu_s$ scales as $\mu_s \sim K_2 n / {d_{\eff}}$ with $C$ and $K_2$ positive constants independent of the system size.

\end{document}